\documentclass[12pt]{article}
\usepackage[reqno]{amsmath}
\usepackage{epsfig}
\usepackage{array}
\usepackage{float}
\usepackage{lscape,graphicx}

\def\ltap{\ \raisebox{-.4ex}{\rlap{$\sim$}} \raisebox{.4ex}{$<$}\ }
\def\gtap{\ \raisebox{-.4ex}{\rlap{$\sim$}} \raisebox{.4ex}{$>$}\ }
\newcommand{\deltaatm}{\mbox{$\Delta  m^2_{\mathrm{A}} $}}
\newcommand{\deltasol}{\mbox{$ \Delta  m^2_{\odot} \ $}}

\newcommand{\betabeta}{\mbox{$(\beta \beta)_{0 \nu}  $}}
\newcommand{\meffnospace}{\mbox{$\left| < \! m  \! > \right|$}}
\newcommand{\meff}{\mbox{$\left|  < \!  m \!  > \right| \ $}}
\newcommand{\meffnhmin}
   {\mbox{$\left|  \langle m  \rangle \right|_{\rm MIN}\ $}}
\newcommand{\meffmin}{\mbox{$\left|  
     < \! m  \! > \right|_{\mbox{}_\mathrm{MIN}} \ $}}
\newcommand{\hbeta}{$\mbox{}^3 {\rm H}$ $\beta$-decay \ }
\newcommand{\eV}{\mbox{$ \  \mathrm{eV} \ $}}

\newcommand{\meffref}{\mbox{$\mu$ }}
\newcommand{\mmin}{\mbox{$m_{\mbox{}_{\rm MIN}} \!\!\!$ }}
\newcommand{\pmns}{\mbox{$ U_{\rm PMNS}$}}
\newcommand{\bea}{\begin{equation}\begin{array}{c}}
\newcommand{\eea}{\end{array}\end{equation}}
\newcommand{\ea}{\end{array}} 

\newcommand{\beq}{\begin{equation}}
\newcommand{\eeq}{\end{equation}}
\newcommand{\bad}{\begin{array}{ccc}}
\newcommand{\dmsol}{\mbox{$\Delta m^2_{\odot}$}}
\newcommand{\mefff}{\mbox{$ < \! m \! > $}}
\newcommand{\ba}{\begin{array}{c}}
\begin{document}
\hfill{Ref. SISSA 78/2007/EP}

\hfill{Ref. IPPP/07/83}

\hfill{Ref. CPT/07/166}

\rightline{hep-ph/xxxxxxx}

\begin{center}
{\bf{\large Majorana Neutrinos, Neutrino Mass Spectrum  and}} 

{\bf{\large the $\meff \sim 10^{-3}$ eV Frontier 
in Neutrinoless Double Beta Decay}}

\vspace{0.4cm}

S. Pascoli$\mbox{}^a$
~and~
S. T. Petcov$\mbox{}^b$
\footnote{Also at: Institute of Nuclear Research and
Nuclear Energy, Bulgarian Academy of Sciences, 1784 Sofia, Bulgaria}

\vspace{0.2cm}
$\mbox{}^a${\em  IPPP, Department of Physics, Durham University, 
Durham, DH1 3LE, United Kingdom\\}

\vspace{0.1cm}
$\mbox{}^b${\em  Scuola Internazionale Superiore di Studi Avanzati, 
I-34014 Trieste, Italy\\
}
{\em Istituto Nazionale di Fisica Nucleare, 
Sezione di Trieste, I-34014 Trieste, Italy\\
}
\end{center}

\begin{abstract}
 
If future neutrino oscillation experiments show 
that the neutrino mass spectrum is with normal ordering, 
$m_1 < m_2 < m_3$, and the searches for neutrinoless 
double beta ($\betabeta$-) decay with sensitivity 
to values of the effective Majorana 
mass $\meff \gtap 10^{-2}$ eV 
give negative results, the next frontier 
in the quest for $\betabeta$-decay 
will correspond to 
$\meff \sim 10^{-3}$ eV. Assuming  
that massive neutrinos are Majorana 
particles and their exchange 
is the dominant mechanism generating 
$\betabeta$-decay, we analise the conditions 
under which $\meff \!\!$, in the case of 
three neutrino mixing and
neutrino mass spectrum 
with normal ordering, would 
satisfy $\meff \geq 0.001$ eV.
We consider the specific cases of 
i) normal hierarchical neutrino mass spectrum, 
ii) of relatively small value of 
the CHOOZ angle $\theta_{13}$ as well as 
iii) the general case of spectrum with normal ordering, 
partial hierarchy and a value of 
$\theta_{13}$ close to the existing 
upper limit. We study the ranges of 
the lightest neutrino mass $m_1$ 
and/or of $\sin^2\theta_{13}$,
for which $\meff \geq 0.001$ eV
and discuss the phenomenological 
implications of such scenarios.
We provide also an estimate 
of $\meff$ when the three neutrino masses and the neutrino 
mixing originate from neutrino mass term of Majorana type 
for the (left-handed) flavour neutrinos and 
$\sum^{3}_{j} m_j U^2_{ej} = 0$, but 
there does not exist a symmetry which forbids 
the $\betabeta$-decay. 

\end{abstract}

\newpage

\section{Introduction}

\vspace{-0.2cm}
\hskip 1cm 
The experiments with solar~\cite{solar,SKsolar,sno}, 
atmospheric~\cite{SKatm}, reactor~\cite{kamland1,kamland}
and accelerator neutrinos~\cite{k2k,MINOS}
have provided during the last several years 
compelling evidence for the existence 
of neutrino oscillations caused by nonzero 
neutrino masses and neutrino mixing. 
The neutrino oscillation data
(see also \cite{LSND,MiniB})
imply the presence of 
3-neutrino mixing in the weak charged 
lepton current
(see, e.g.~\cite{STPNu04}):
\begin{equation}
\nu_{l \mathrm{L}}  = \sum_{j=1}^{3} U_{l j} \, \nu_{j \mathrm{L}},~~
l  = e,\mu,\tau,
\label{3numix}
\end{equation}
%
where $\nu_{l\mathrm{L}}$ are the flavour neutrino fields, 
$\nu_{j \mathrm{L}}$ is the field of neutrino 
$\nu_j$ having a mass $m_j$ and $U$ is the
Pontecorvo--Maki--Nakagawa--Sakata (PMNS) mixing 
matrix~\cite{BPont57}, $U \equiv \pmns$. 

   In spite of the remarkable progress 
made, first, in demonstrating experimentally
the existence of neutrino oscillations
and, second, in determining 
the pattern  of neutrino mixing and the 
values of the two neutrino mass squared 
differences, 
responsible for the solar and 
atmospheric neutrino oscillations, 
our knowledge in what concerns
most of the basic aspects 
of neutrino mixing is very 
limited at present (see, e.g. \cite{STPNu04}). 
We still do not know   
i) what the nature 
of neutrinos with definite mass is - 
Dirac or Majorana,
ii) what type of spectrum neutrino 
masses obey, iii) what the absolute scale of neutrino masses is, 
iv) whether the CP-symmetry is violated in the lepton sector 
by the neutrino mixing matrix $\pmns$, 
v) what the value of the CHOOZ angle is - being the smallest 
mixing angle in the PMNS matrix, it controls 
(together with the Dirac CP-violating phase) 
the magnitude of CP-violation effects in neutrino 
oscillations,
vi) whether the observed patterns of neutrino mixing 
is related to the existence of a new  
symmetry in Nature, etc.

  Establishing whether the neutrinos with definite mass
$\nu_j$ are Dirac fermions possessing distinct antiparticles, 
or Majorana fermions, i.e. spin 1/2 particles that 
are identical with their antiparticles, is 
of fundamental importance for making progress in 
our understanding of the origin of
neutrino masses and mixing and
of the symmetries governing the lepton sector
of particle interactions (see, e.g. \cite{STPNu04}).
It is well-known that 
the presence of massive Dirac 
neutrinos is associated 
with the existence of 
a conserved additive lepton number, 
which can be, e.g. the total
lepton charge $L = L_e + L_{\mu} + L_{\tau}$. 
If the particle interactions do not 
conserve any lepton charge,
the massive neutrinos $\nu_j$ will be Majorana 
fermions (see, e.g.~\cite{BiPet87}). 

  The only feasible experiments 
having the potential of establishing the
Majorana nature of massive neutrinos 
at present are the
$\betabeta$-decay experiments searching 
for the process $(A,Z) \rightarrow (A,Z+2) + e^- + e^-$
(for reviews see, e.g. 
\cite{BiPet87,Morales02,APSbb0nu,STPFocusNu04}).
The observation of \betabeta-decay
and the measurement of the corresponding 
half-life with sufficient accuracy,
would not only be a proof that the total 
lepton charge is not conserved, 
but might provide also unique 
information on the i) type of neutrino 
mass spectrum 
\cite{PPSNO2bb,BPP1} (see also \cite{PPW,PPRSNO2bb}),
ii) absolute scale of neutrino 
masses
(see, e.g. \cite{PPW}), 
and iii) Majorana CP-violating (CPV) phases 
\cite{BGKP96,BPP1,PPR1,PPSchw05}
(see also the related discussions 
in, e.g. \cite{BGGKP99,WR00,LMWR05}).

  Under the assumptions 
of 3-$\nu$ mixing,
of massive neutrinos $\nu_j$ being
Majorana particles, and
of \betabeta-decay generated
only by the (V-A) charged current 
weak interaction via the exchange of
the three Majorana neutrinos  
$\nu_j$ having masses
$m_j \ltap$ few MeV,
the $\betabeta$-decay amplitude  
has the form (see, e.g. \cite{BiPet87,BPP1}): 
$A\betabeta \cong \mefff~M$, where 
$M$ is the corresponding 
nuclear matrix element (NME) which does not 
depend on the neutrino mixing parameters, and
$\mefff$ is the \betabeta-decay effective Majorana mass,
\begin{equation}
\meff=\left| m_1\,  
|U_{\mathrm{e} 1}|^2 
+ m_2\, |U_{\mathrm{e} 2}|^2\, e^{i\alpha_{21}}
+ m_3\, |U_{\mathrm{e} 3}|^2\, e^{i\alpha_{31}} 
\right|\,.
\label{effmass2}
\end{equation}
%
\noindent Here 
$|U_{\mathrm{e}j}|$, $j=1,2,3$, 
are the absolute values of the elements of the
first raw of the PMNS mixing matrix,
$|U_{\mathrm{e}1}| = c_{12}c_{13}$,
$|U_{\mathrm{e}2}| = s_{12}c_{13}$, 
$|U_{\mathrm{e}3}| = s_{13}$,
$c_{ij} \equiv \cos\theta_{ij}$, 
$s_{ij} \equiv \sin\theta_{ij}$,
$\theta_{12} \equiv \theta_{\odot}$, 
$\theta_{23} \equiv \theta_{\rm A}$ 
and $\theta_{13}$ being the solar neutrino, 
atmospheric neutrino and 
CHOOZ mixing angles in the standard parametrisation 
of $\pmns$ (see, e.g. \cite{BPP1}),
and $\alpha_{21}$, $\alpha_{31}$
are the two Majorana CP-violation phases in $\pmns$
\cite{BHP80,Doi81SchValle80}.

\indent The experimental searches 
for $\betabeta$-decay  
have a long history \cite{Morales02}. 
The best sensitivity was achieved in the
Heidelberg-Moscow $^{76}$Ge experiment 
\cite{HMGe76}: $\meff<$(0.35 - 1.05) eV (90\% C.L.),
where a factor of 3 uncertainty in the relevant NME 
(see, e.g. \cite{FesSimVogel03})
is taken into account.
The IGEX collaboration has obtained
\cite{IGEX00}: $\meff < (0.33 - 1.35)$ eV (90\% C.L.).
A positive signal at $> 3\sigma$,
corresponding to
$\meff = (0.1 - 0.9)~{\rm eV}$,
is claimed to be observed in \cite{Klap04}. 
Two experiments, NEMO3 (with $^{100}$Mo and 
$^{82}$Se) \cite{NEMO3}
and CUORICINO (with $^{130}$Te) \cite{CUORI},
designed to reach a sensitivity to 
$\meff\sim (0.2-0.3)$ eV, 
set the limits: $\meff < (0.7 - 1.2)$ eV \cite{NEMO3} 
and $\meff< (0.2 - 0.9)$ eV
\cite{CUORI} (90\% C.L.),
where estimated uncertainties in the NME
are accounted for. Most importantly,
a large number of projects aim at a sensitivity to 
$\meff \sim$(0.01--0.05) eV 
\cite{bb0nu}:
CUORE ($^{130}$Te), GERDA ($^{76}$Ge),
SuperNEMO, EXO ($^{136}$Xe), MAJORANA ($^{76}$Ge),
MOON ($^{100}$Mo), COBRA ($^{116}$Cd), 
XMASS ($^{136}$Xe), CANDLES ($^{48}$Ca), etc. 
These experiments, in particular, will 
test the positive result claimed in \cite{Klap04}.

  The predicted value of \meff{} 
depends strongly on the 
type of $\nu-$mass spectrum 
\cite{PPSNO2bb,BPP1}, more precisely, 
on the type of hierarchy neutrino 
masses obey. Let us recall that
the neutrino mass spectrum 
(in a standardly used convention)
can be with {\it normal ordering}, 
$m_1 < m_2 < m_3$, 
or with {\it inverted ordering}, 
$m_3 < m_1 < m_2$. 
The first corresponds to $\deltaatm \equiv \Delta m^2_{31} > 0$,
$|\deltaatm| \sim (0.05)^2~{\rm eV^2}$ being the neutrino 
mass squared difference 
responsible for the (dominant) 
atmospheric neutrino oscillations;
the second is realised if 
$\deltaatm \equiv \Delta m^2_{32} < 0$.
Depending on the 
${\rm sgn}(\deltaatm)$ 
and the value of the lightest 
neutrino mass, i.e., the absolute 
neutrino 
mass scale, ${\rm min}(m_j) \equiv \mmin$, the 
neutrino mass spectrum can be \\ 
i) {\it Normal Hierarchical} (NH): $m_1 \ll m_2 < m_3$,
$m_2 \cong (\dmsol)^ {1\over{2}}$, 
$m_3 \cong (\deltaatm)^{1\over{2}}$, 
$\dmsol \equiv \Delta m^2_{21} \sim 0.009$ eV being 
the neutrino mass squared difference
driving the solar 
$\nu_e$ oscillations;\\
ii) {\it Inverted Hierarchical} (IH): $m_3 \ll m_1 < m_2$,
with $m_{1,2} \cong |\deltaatm|^{1\over{2}}$, 
$\dmsol = \Delta m^2_{21}$;\\
iii) {\it Quasi-Degenerate} (QD): $m_1 \cong m_2 \cong m_3 \cong m_0$,
$m_j^2 \gg |\deltaatm|$, $m_0 \gtap 0.10$~eV.

\indent  The existence of significant 
and robust lower bounds on $\meff$
in the cases of IH and QD spectra 
\cite{PPSNO2bb} (see also \cite{PPW}), 
given respectively 
\footnote{Up to small corrections 
we have in the cases of two spectra \cite{PPSNO2bb}:
$\meff \gtap \deltaatm \cos2\theta_{\odot}$ (IH) and
$\meff \gtap m_0 \cos2\theta_{\odot}$ (QD).
The possibility of 
$\cos2\theta_{\odot} = 0$
is ruled out at $\sim 6\sigma$ 
by the existing data \cite{BCGPRKL2,TSchwNuFact07}, 
which also imply that
$\cos 2\theta_\odot \gtap 0.26$ 
at 2$\sigma$~\cite{TSchwNuFact07}. We also have 
$\deltaatm \gtap 2.0\times 10^{-3}~{\rm eV^2}$ 
at 3$\sigma$(see further).}
by $\meff \gtap 0.01$ eV and $\meff \gtap 0.03$ eV,
which lie either partially (IH spectrum) or completely
(QD spectrum) within the range of sensitivity of 
the next generation of \betabeta-decay experiments,
is one of the most important features of
the predictions of $\meff\!\!$. 
At the same time we have $\meff \ltap 5\times 10^{-3}$ eV
in the case of NH spectrum \cite{PPSchw05}.
The fact that  $max(\meff)$ in the case of NH 
spectrum is considerably smaller than
$min(\meff)$ for the IH and QD spectrum
opens the possibility of obtaining
information about the type of 
$\nu$-mass spectrum from a 
measurement of $\meff \neq 0$
~\cite{PPSNO2bb}.
More specifically, a positive result 
in the future generation
of \betabeta-decay experiments with $\meff > 0.01$ eV
would imply that the NH spectrum is strongly 
disfavored (if not excluded).
For $\deltaatm > 0$, 
such a result would mean that the 
neutrino mass spectrum is 
with normal ordering, but is 
not hierarchical. If $\deltaatm < 0$, 
the neutrino mass spectrum would be 
either IH or QD. 

  If the future \betabeta-decay experiments 
show that $\meff < 0.01$ eV, both the IH and
the QD spectrum will be ruled out for massive 
Majorana neutrinos. If in addition it is 
established in neutrino oscillation 
experiments that the neutrino mass spectrum is 
with {\it inverted ordering}, i.e. 
that $\deltaatm < 0$,
one would be led to conclude that 
either the massive neutrinos $\nu_j$ 
are Dirac fermions, or that 
$\nu_j$ are Majorana particles
but there are additional contributions to the 
\betabeta-decay amplitude which 
interfere distructively 
with that due to the exchange of 
light massive Majorana neutrinos. 
However, if $\deltaatm$ is determined to be positive
in neutrino oscillation experiments, 
the upper limit $\meff < 0.01$ eV 
would be perfectly compatible with 
massive Majorana neutrinos
possessing NH mass spectrum, 
or mass spectrum with normal ordering but
partial hierarchy, and the quest for $\meff$ 
would still be open.

\indent  If indeed in the next generation of 
\betabeta-decay experiments it is 
found that $\meff < 0.01$ eV, 
while the neutrino oscillation experiments 
show that  $\deltaatm > 0$, 
the next frontier in the searches for 
$\betabeta-$decay would most probably 
correspond to values of $\meff \sim 0.001$ eV.
Taking $\meff = 0.001$ eV as a reference value,
we investigate in the present article
the conditions under which $\meff$ in the 
case of neutrino mass spectrum with normal 
ordering would be guaranteed to satisfy 
$\meff \gtap 0.001$ eV.
We consider the specific cases of 
i) normal hierarchical neutrino mass spectrum, 
ii) of relatively small value of 
the CHOOZ angle $\theta_{13}$ as well as 
iii) the general case of spectrum with normal ordering, 
partial hierarchy and a value of 
$\theta_{13}$ close to the existing 
upper limit. We study the ranges of 
the lightest neutrino mass $m_1$ 
and/or of $\sin^2\theta_{13}$,
for which $\meff \gtap 0.001$ eV
and discuss the phenomenological 
implications of such scenarios.

 In the present analysis we do not 
include the effect of the uncertainty related 
to the imprecise knowledge of the 
$\betabeta-$decay nuclear matrix elements
(see, e.g. \cite{FesSimVogel03}).
We hope that by the time it will become 
clear whether the searches for 
$\betabeta-$decay will require 
a sensitivity to values of 
$\meff < 0.01$ eV, the problem of 
sufficiently precise
calculation of the $\betabeta-$decay 
nuclear matrix elements  will be resolved 
\footnote{Encouraging results,
in what regards the problem of 
calculation of the NME, were 
reported in \cite{FesSimVogel03}.
A possible test of the NME calculations 
is discussed in \cite{NMEBiPet04}.
Let us note that nuclear matrix 
elements uncertainties do not affect 
the predictions for the effective Majorana 
mass parameter directly, but induce a spread on 
the values of the $\betabeta$-decay
half-life times which correspond to the 
predicted values of $\meff$.  
Conversely, if a measurement of the half-life time
is performed or a stringent bound is obtained, 
they would affect the experimentally 
determined value of $\meff$ and the 
constraints following from the latter.}.

  The paper is organised as follows.
In Section 2 we present predictions for $\meff$
using the present 2$\sigma$ experimentally allowed 
ranges of values of the neutrino oscillation parameters 
and future prospective uncertainties in their values. 
In Section 3 we analise the conditions under 
which $\meff$ in the case of $\nu$ 
mass spectrum with normal ordering would 
be guaranteed to satisfy $\meff \gtap 0.001$ eV.
We consider the cases of i) normal hierarchical spectrum, 
ii) small $\theta_{13}$, and 
ii) spectrum with partial hierarchy. 
In Section 4  we give an estimate 
of $\meff$ when the three $\nu$ masses and the neutrino 
mixing originate from neutrino mass term of Majorana type 
for the (left-handed) flavour neutrinos and 
$\sum^{3}_{j} m_j U^2_{ej} = 0$, but 
$\betabeta$-decay is allowed.
Section 5 contains the conclusions of the present analysis.

%
\section{Neutrino Oscillation Data and Predictions for \meff}
%

\vspace{-0.2cm}
The existing neutrino oscillation data
allow us to determine 
the parameters which drive the 
solar neutrino and the 
dominant atmospheric neutrino 
oscillations,
$\dmsol = \Delta m^2_{21}$, 
$\sin^2\theta_{12} \equiv \sin^2\theta_{\odot}$, and
$|\deltaatm| = |\Delta m^2_{31}| \cong |\Delta m^2_{32}|$, 
$\sin^22\theta_{23}$, with a relatively 
good precision, and to obtain
rather stringent limits on the CHOOZ angle \cite{chooz}
$\theta_{13}$ 
(see, e.g.~\cite{BCGPRKL2,TSchwNuFact07}).
The best fit values and the 2$\sigma$ allowed ranges of 
$|\deltaatm|$, 
$\dmsol$ and $\sin^2\theta_{\odot}$
read~\cite{TSchwNuFact07}:
\begin{eqnarray}
\label{deltaatmvalues}
(|\deltaatm|)_{\rm BF} = 2.4 \times 10^{-3} \ \eV^2, 
& 2.1  \times 10^{-3} \ \eV^2 \leq |\deltaatm| 
\leq 2.7  \times 10^{-3} \ \eV^2 ,  \\
\label{deltasolvalues}
(\deltasol)_{\rm BF} = 7.6 \times 10^{-5} \ \eV^2, 
& 7.3 \times 10^{-5} \ \eV^2 \leq \deltasol \leq 8.1 \times 10^{-5} \ \eV^2, \\
\label{sinsolvalues}
(\sin^2 \theta_\odot)_{\rm BF} = 0.32,
& 0.28 \leq \sin^2 \theta_\odot \leq 0.37\,.
\end{eqnarray}
%
A combined 3-$\nu$ oscillation analysis of the global 
neutrino oscillation data
gives~\cite{TSchwNuFact07}
\vspace{-0.1cm}
\beq
\sin^2\theta_{13} < 0.033~(0.050)
\quad\mbox{at}\quad 2\sigma \, (3\sigma)~.
\label{th13}
\eeq
%
  The existing data allow a determination of
$\dmsol$, $\sin^2\theta_{\odot}$
and $|\deltaatm|$ at 3$\sigma$ with an 
error of approximately 8\%, 22\%, 
and 17\%, respectively~\cite{TSchwNuFact07}.
Future oscillation experiments 
will improve considerably the 
precision on these basic parameters:
the indicated 3$\sigma$ errors could be reduced to 
4\%, 12\%~\cite{SKGdCP04,reactoaccfuture} and better than
5\%~\cite{reactoaccfuture,Donini:2005db,Blondel:2006su}
(see also the discussion in \cite{STPNu04,PPSchw05}
and the references quoted therein), and even 
to $\sim 1\%$ for \deltaatm \cite{Bross:2007ts}.
``Near'' future experiments with reactor $\bar{\nu_e}$ 
can improve the current sensitivity to the value 
of  $\sin^2\theta_{13}$ by a factor of (5-10) 
(see, e.g. \cite{Reacth13}), while future long baseline experiments
will aim at measuring values of $\sin^2\theta_{13}$ as small
as $10^{-4}$--$10^{-3}$~(see, e.g. \cite{reactoaccfuture,Blondel:2006su}).

The type of neutrino mass hierarchy, i.e.
${\rm sgn}(\deltaatm)$, can be determined 
by studying oscillations of neutrinos and
antineutrinos, say, $\nu_{\mu} \leftrightarrow \nu_e$
and $\bar{\nu}_{\mu} \leftrightarrow \bar{\nu}_e$,
in which matter effects are sufficiently large.
This can be done in long base-line 
$\nu$-oscillation experiments 
(see, e.g. \cite{Future,reactoaccfuture,Blondel:2006su}). 
If $\sin^22\theta_{13}\gtap 0.05$
and $\sin^2\theta_{23}\gtap 0.50$,
information on ${\rm sgn}(\Delta m^2_{31})$
might be obtained in atmospheric 
neutrino experiments by investigating 
the effects of the subdominant transitions
$\nu_{\mu(e)} \rightarrow \nu_{e(\mu)}$
and $\bar{\nu}_{\mu(e)} \rightarrow \bar{\nu}_{e(\mu)}$ 
of atmospheric  neutrinos which traverse 
the Earth \cite{JBSP203}. 
For $\nu_{\mu(e)}$ ({\it or} $\bar{\nu}_{\mu(e)}$) 
crossing the Earth core, new type of resonance-like
enhancement of the indicated transitions
takes place due to the {\it (Earth) mantle-core
constructive interference effect
(neutrino oscillation length resonance (NOLR))} 
\cite{SP3198}~
\footnote{As a consequence of this effect
the corresponding 
$\nu_{\mu(e)}$ ({\it or} $\bar{\nu}_{\mu(e)}$)
transition probabilities can be maximal \cite{106107}
(for the precise conditions
of the mantle-core (NOLR) enhancement
see \cite{SP3198,106107}).
Let us note that the Earth mantle-core (NOLR) enhancement of
neutrino transitions differs \cite{SP3198} from the MSW 
one. It also differs \cite{SP3198,106107} from the 
parametric resonance mechanisms of enhancement 
discussed in the articles \cite{Param86}.
}. 
For $\Delta m^2_{31}> 0$, the neutrino transitions
$\nu_{\mu(e)} \rightarrow \nu_{e(\mu)}$
are enhanced, while for $\Delta m^2_{31}< 0$
the enhancement of antineutrino transitions
$\bar{\nu}_{\mu(e)} \rightarrow \bar{\nu}_{e(\mu)}$
takes place, which might allow 
to determine ${\rm sgn}(\Delta m^2_{31})$.
If $\sin^2\theta_{13}$ is sufficiently large, 
the sign of $\deltaatm$ can 
also be determined by studying 
the oscillations of reactor $\bar{\nu}_e$ 
on distances of $\sim (20 -40)$ km \cite{PiaiP0103}. 
An experiment with reactor $\bar{\nu}_e$,
which, in particular, might have the 
the capabilities to measure
${\rm sgn}(\deltaatm)$,
was proposed recently in \cite{Hano}.
According to \cite{Hano}, 
this experiment can provide 
a determination of $|\deltaatm|$
with an uncertainty of 
$(3 - 4)\%$ at 3$\sigma$.

  As is well-known, 
neutrino oscillations are not sensitive to
the absolute scale of neutrino masses.
Information on the absolute 
neutrino mass scale 
can be derived in \hbeta experiments 
\cite{Fermi34,MoscowH3,MainzKATRIN}
and from cosmological and astrophysical data. 
The most stringent upper 
bounds on the $\bar{\nu}_e$ mass 
were obtained in the Troitzk~\cite{MoscowH3} 
and Mainz~\cite{MainzKATRIN} experiments: 
\vspace{-0.10cm}
\beq
m_{\bar{\nu}_e} < 2.3 \  \mathrm{eV}~~~\mbox{at}~95\%~\mathrm{C.L.} 
\label{H3beta}
\eeq
%
\noindent We have $m_{\bar{\nu}_e} \cong m_{1,2,3}$
in the case of the QD $\nu$-mass spectrum.
The KATRIN experiment~\cite{MainzKATRIN}
is planned to reach a sensitivity  
of  $m_{\bar{\nu}_e} \sim 0.20$~eV,
i.e. it will probe the region of the QD 
spectrum. 
  Information on the type of neutrino mass spectrum
can also be obtained in $\beta$-decay experiments 
having a sensitivity to neutrino masses \cite{BMP06} 
$\sim \sqrt{|\deltaatm|}\cong 5\times 10^{-2}$ eV
(i.e. by a factor of $\sim 4$ better sensitivity than 
KATRIN \cite{MainzKATRIN}).

The 
CMB data of the WMAP experiment \cite{WMAPnu}, 
combined with data from large scale structure surveys 
(2dFGRS, SDSS), lead to the following
upper limit on the sum of 
neutrino masses 
(see, e.g. \cite{Hann06}):
\beq
\sum_{j} m_{j} \equiv \Sigma < (0.4 \mbox{--} 1.7)~
{\rm eV~~~\mbox{at}~95\%~C.L.} 
\label{WMAP}
\eeq
%
Data on weak lensing of galaxies, 
combined with data from the WMAP and PLANCK
experiments, may allow $\Sigma$ to be determined 
with an uncertainty of 
$ \sim 0.04$~eV \cite{Hann06,Hu99}.

   It proves convenient to express   
\cite{SPAS94} the three neutrino masses
in terms of $\dmsol$ and $\deltaatm$, measured 
in neutrino oscillation experiments,
and the absolute neutrino  mass scale
determined by ${\rm min}(m_j)\equiv \mmin$  
\footnote{For a detailed discussion of 
the relevant formalism 
see, e.g. \cite{BPP1,STPFocusNu04}.}.
In both cases of $\nu$-mass spectrum 
with normal and inverted ordering one has
(in the convention we use):
$\dmsol=\Delta m_{21}^2 > 0$, 
$m_2=(m_1^2 + \dmsol)^{\frac{1}{2}}$.
For normal ordering, $\mmin \equiv m_1$, 
$\deltaatm=\Delta m_{31}^2 > 0$
and $m_3=(m_1^2 + \deltaatm)^{\frac{1}{2}}$,
while if the spectrum is with inverted ordering,
$\mmin=m_3$, $\deltaatm=\Delta m_{32}^2 < 0$ and 
$m_1=(m_3^2 + |\deltaatm| - \dmsol)^{\frac{1}{2}}$. 
For the elements of the PMNS matrix $|U_{\mathrm{e} j}|^2$, 
$j=1,2,3$,
as we have already indicated, 
the following relations hold:
$|U_{\mathrm{e} 1}|^2 = \cos^2\theta_{\odot} (1 - \sin^2 \theta_{13})$, 
$|U_{\mathrm{e} 2}|^2 = \sin^2\theta_{\odot} (1 - \sin^2 \theta_{13})$,
and  $|U_{\mathrm{e} 3}|^2 \equiv \sin^2\theta_{13}$.
Thus, given $|\deltaatm|$, $\dmsol$, $\theta_{\odot}$ 
and $\theta_{13}$, $\meff$ depends 
on the lightest neutrino mass 
(absolute neutrino mass scale), \mmin, 
the two Majorana phases $\alpha_{21}$ and $\alpha_{31}$, 
present in the PMNS matrix 
and on the type of neutrino mass spectrum (see, e.g. \cite{BPP1}). 
For neutrino mass spectrum with normal ordering we have
\begin{eqnarray} 
\meff &=& \left| \mmin \cos^2 \theta_\odot ( 1 - \sin^2 \theta_{13}) 
+ \sqrt{\mmin^{\!\!\!\!\!\!\!\!2} \  + \deltasol} \sin^2 \theta_\odot ( 1 - \sin^2 \theta_{13}) 
e^{i \alpha_{21}} \right.
\nonumber \\ [0.25cm]
& & \left. + \sqrt{\mmin^{\!\!\!\!\!\!\!\!2} \  + 
\deltaatm} \sin^2 \theta_{13} e^{i \alpha_{31}} \right|\,,~~~
\mmin \equiv m_1\,.
\label{meffcompleteNH}
\end{eqnarray}
%
For spectrum with inverted ordering a  
different expression is valid \cite{BGKP96,BPP1}:
\begin{eqnarray} 
\meff &=&  \left|\sqrt{\mmin^{\!\!\!\!\!\!\!\!2} \  + |\deltaatm| - \deltasol}\, 
\cos^2 \theta_\odot ( 1 - \sin^2 \theta_{13}) \right. 
\nonumber \\ [0.25cm]
   &&\left. + \, \sqrt{\mmin^{\!\!\!\!\!\!\!\!2} \  + |\deltaatm|} 
\sin^2 \theta_\odot ( 1 - \sin^2 \theta_{13}) 
e^{i \alpha_{21}} +\mmin \sin^2 \theta_{13} 
e^{i \alpha_{31}} \right|\,
\label{meffcompleteIH}
\\[0.25cm]
 &\cong & \sqrt{\mmin^{\!\!\!\!\!\!\!\!2} \  + |\deltaatm|}\,
\left |\cos^2 \theta_\odot + 
\sin^2 \theta_\odot\, e^{i \alpha_{21}}\right |\,
( 1 - \sin^2 \theta_{13})\,,~\mmin \equiv m_3\,.
\label{meffapprIH}
\end{eqnarray}
%
In Eq. (\ref{meffapprIH}) we have neglected  
$\deltasol$ with respect to $(\mmin^{\!\!\!\!\!\!\!\!2} \  \ + |\deltaatm|)$ 
and the term $\mmin \sin^2 \theta_{13}$. According to 
the existing data, we have 
$\deltasol/(\mmin^{\!\!\!\!\!\!\!\!2} \  \  + |\deltaatm|) \ltap 0.032$, and 
$\mmin \sin^2 \theta_{13} \ll 
(\mmin^{\!\!\!\!\!\!\!\!2} \  + 
|\deltaatm|)^{\frac{1}{2}}\,\cos2 \theta_\odot$.
Actually, the term $\mmin \sin^2 \theta_{13}$ can always 
be neglected 
provided $\sin^2 \theta_{13} \ll \cos2 \theta_\odot$.
The expression for $\meff$ in the case of 
IH spectrum follows from Eq. (\ref{meffapprIH}) 
if $\mmin^{\!\!\!\!\!\!\!\!2}~\ll |\deltaatm|$ and
$\mmin^{\!\!\!\!\!\!\!\!2} \ $~ is neglected  with respect to $|\deltaatm|$.
For the QD spectrum we get:
\begin{eqnarray} 
\meff &=& m_0\, \left| 
\left ( \cos^2 \theta_\odot + \sin^2 \theta_\odot\, e^{i \alpha_{21}}\right ) 
( 1 - \sin^2 \theta_{13}) + \sin^2 \theta_{13}\, 
e^{i \alpha_{31}} \right|\, ,
\label{meffcompleteQD}
\\[0.25cm]
&\cong& m_0\,\left | \cos^2 \theta_\odot + 
\sin^2 \theta_\odot\, e^{i \alpha_{21}}\right |\,
( 1 - \sin^2 \theta_{13})\,,
\label{meffapprQD}
\end{eqnarray}
%
where $m_0\equiv \mmin$, $m_1\cong m_2 \cong m_3$.
Evidently, as long as $\sin^2 \theta_{13} \ll \cos2 \theta_\odot$,
the terms $\propto \sin^2\theta_{13}$ in $\meff$ 
play an insignificant role in the cases 
of neutrino mass spectrum with inverted ordering 
(i.e. $\deltaatm < 0$), or of QD type 
(for any  ${\rm sgn}(\deltaatm)$).
In what concerns the spectrum with normal ordering, 
the term $\sqrt{\mmin^{\!\!\!\!\!\!\!\!2} \  + \deltaatm} \sin^2 \theta_{13}$
can be crucial for determining 
the magnitude of $\meff$ 
if massive neutrinos are not QD, i.e. if
$\mmin^{\!\!\!\!\!\!\!\!2} \  \ltap \deltaatm$, and 
$\sin^2 \theta_{13}$ is sufficiently large 
(see further).

  If CP-invariance holds, we have \cite{LW81} 
$\alpha_{21} = k\pi$ and $\alpha_{31} = k'\pi$,
$k,k'=0,1,2,...$. In the case of CP-invariance 
the phase factors 
\begin{equation}
\eta_{21} \equiv e^{i\alpha_{21}}= \pm 1\,,~ 
\eta_{31}\equiv e^{i\alpha_{31}} = \pm 1\,,~
\eta_{32}\equiv e^{i\alpha_{32}} = \pm 1\,,~   
\label{eta2131}
\end{equation}
%
\noindent as is well-known, have a simple physical 
interpretation \cite{LW81,BiPet87}:
$\eta_{ik}$ is the relative CP-parity 
of Majorana neutrinos $\nu_{i}$ and $\nu_k$.
Obviously, $\meff$ 
depends strongly on the Majorana CPV phase(s):
the CP-conserving values of 
$\alpha_{21} = 0,\pm\pi$ determine, 
for instance, the range of 
possible values of $\meff$ in the 
cases of IH and QD spectrum.

  We recall that the neutrino oscillation 
experiments are insensitive
to the two Majorana CP-violation phases 
in the PMNS matrix \cite{BHP80,Lang87} 
-- the latter do not enter into the 
expressions for the probabilities of 
flavour neutrino oscillations.
It is interesting to note, however, 
that in addition of playing an
important role in the predictions for 
$\meff$ and, correspondingly, of the 
$\betabeta-$decay half-life, 
the Majorana phase(s) in $\pmns$
can provide the CP-violation necessary 
for the generation of the baryon 
asymmetry of the Universe
\cite{PPRio106,MPST07} (see also \cite{SPTSMor06}).
The Majorana phases 
$\alpha_{21}$ and $\alpha_{32}$
can also affect significantly 
the predictions for the 
rates of (LFV) decays $\mu \rightarrow e + \gamma$,
$\tau \rightarrow \mu + \gamma$, etc.
in a large class of supersymmetric theories
with see-saw mechanism of $\nu$-mass generation 
\cite{PPY03}.

  First, we will update the 
predictions for \meff as a function of $\mmin$,
using as input the 2$\sigma$ ranges of values of 
$\deltaatm$, $\deltasol\!$, $\sin^2 \theta_\odot$ and $\sin^2 \theta_{13}$, 
obtained from the latest available set 
of neutrino oscillation data 
(see Eqs.~(\ref{deltaatmvalues}), (\ref{deltasolvalues}) and 
(\ref{sinsolvalues})).
Since $\alpha_{21}$ and $\alpha_{31}$ 
cannot be determined in independent experiments,
we treat them as free parameters 
taking values $0 \leq \alpha_{21,31} \leq 2 \pi$.
The results of this analysis are shown in Fig.~1.
\begin{figure}[p]
 \label{fig:meff95CL}
{\epsfig{file=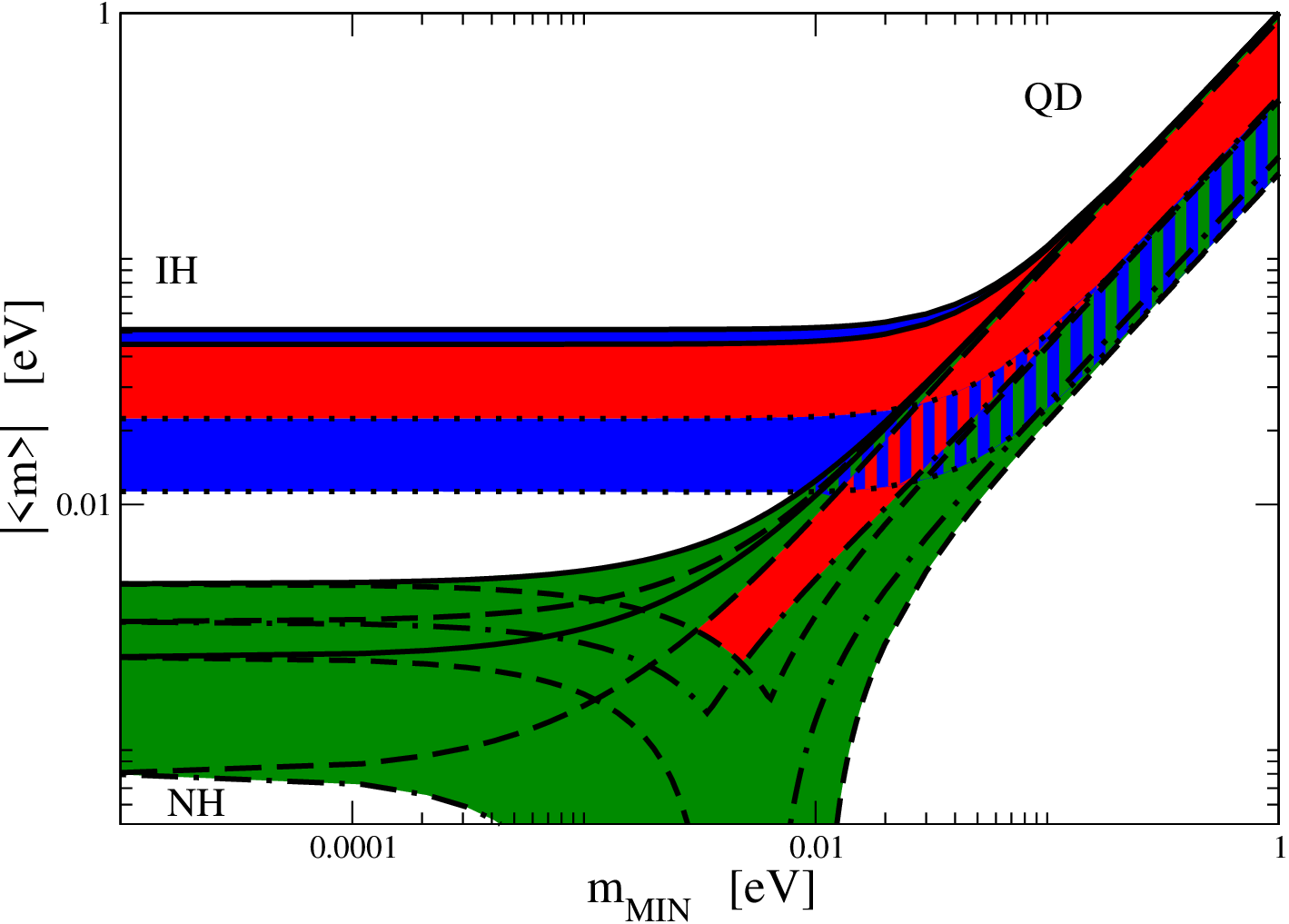, height=10cm, width=16cm}
}
\caption{ The predicted value of $\meff$ as a function of \mmin, 
obtained using the 2$\sigma$ allowed ranges of 
\deltaatm, $\deltasol \!\!$, $\sin^2 \theta_\odot$ 
and $\sin^2 \theta_{13}$.
   For the NH and QD (and interpolating) spectra, the green regions within
   the black lines of a given type (solid, short-dashed, long-dashed,
   dash-dotted) correspond to the four different sets of CP-conserving
   values of the two phases $\alpha_{21}$ and $\alpha_{31}$, and thus
   to the four possible combinations of the relative CP parities
   ($\eta_{21},\eta_{31}$) of neutrinos $\nu_{1,2}$ and $\nu_{1,3}$:
   $(+1,+1)$ solid, $(-1,-1)$ short-dashed, $(+1,-1)$ long-dashed, and
   $(-1,+1)$ dash-dotted lines.  For the IH spectrum, the blue regions
   delimited by the black solid (dotted) lines correspond to
   $\eta_{21} = + 1$ ($\eta_{21} = -1$), independently of $\eta_{31}$.
   The regions shown in red correspond to violation of
   CP-symmetry.
}
\end{figure}
%

We report in Table~1 the maximal and minimal 
values of $\meff$ for the normal hierarchical 
(NH) spectrum, $m_1 \ll m_2 < m_3$,
for the inverted hierarchical 
(IH) spectrum, $m_3 \ll m_1 \simeq m_2$,
and for the quasi-degenerate spectrum (QD),
$m_1 \simeq m_2 \simeq m_3 \geq 0.2$~eV.
\begin{table}[ht]
\begin{center}
\begin{tabular}{|c|c|c|c|c|} 
\hline
\rule{0pt}{0.5cm} 
$\meff_{\rm min}^{\rm NH}$ &$\meff_{\rm max}^{\rm NH}$ 
& $\meff_{\rm min}^{\rm IH}$ &  $\meff_{\rm max}^{\rm IH}$ &
$\meff_{\rm min}^{\rm QD} $  
\\ \hline \hline
0.7  & 4.8 &  11.3    &  51.5   &  44.2 \\ \hline
\end{tabular}
\caption{\label{tab:bf} 
The maximal values of  
$\meff$ (in units of meV)
for the NH  and IH spectra, and the minimal values of 
$\meff$ (in units of meV) for the NH, IH and QD spectra,
obtained using the 2$\sigma$ allowed values 
of the neutrino oscillation parameters.
The results for the NH and IH spectra 
are for $\mmin = 10^{-4}$ eV, while
those for the QD spectrum correspond to 
$m_{\mbox{}_{\mathrm{MIN}}} = 0.2$ eV\@.
}
\end{center}
\end{table}
%

\begin{figure}
 \label{fig:meffbf}
{\epsfig{file=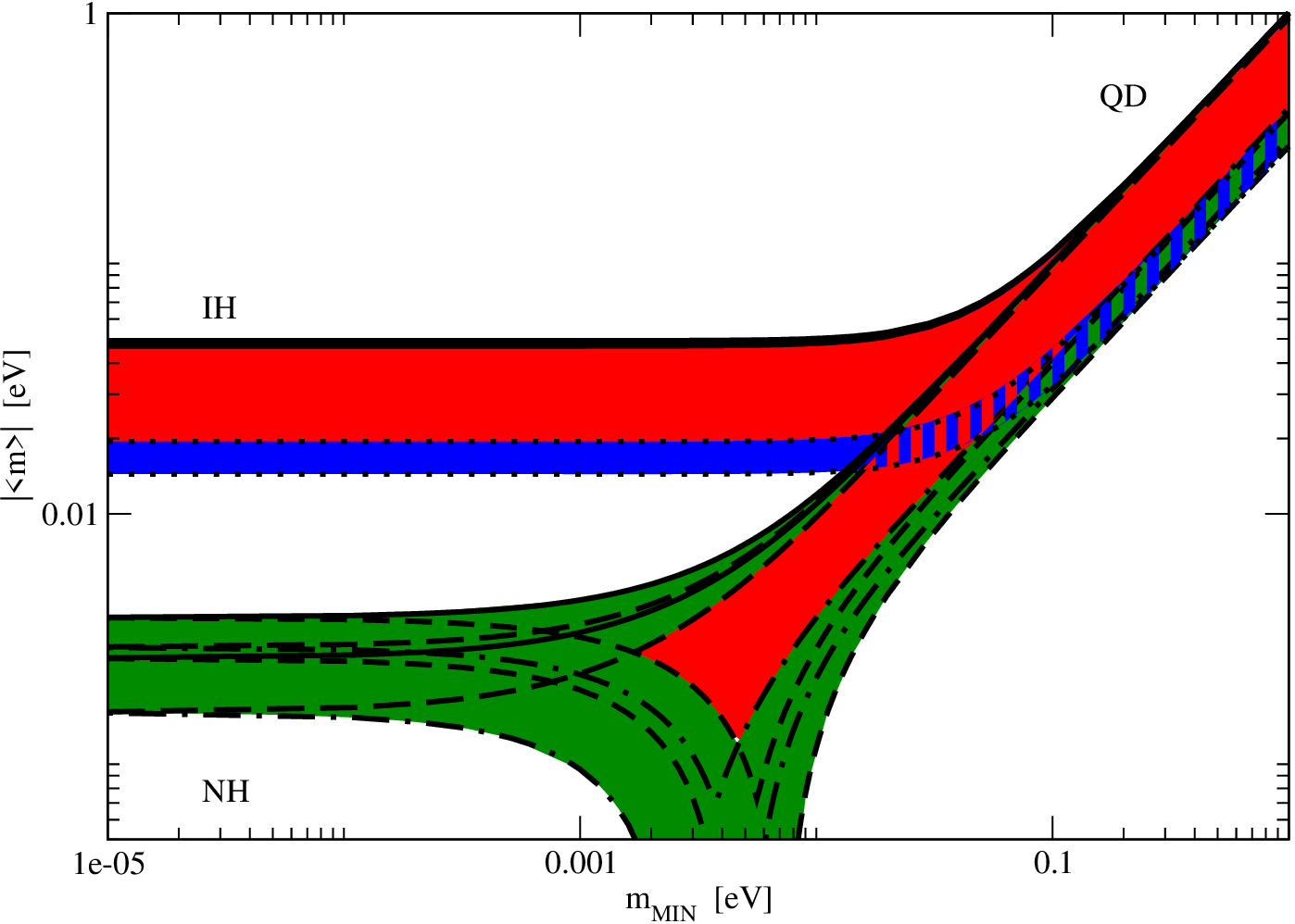, height=10cm, width=16cm}
}
\caption{The predicted value of $\meff$ (including a prospective
   2$\sigma$ uncertainty) as a function of \mmin~ for
   $\sin^2\theta_{13} = 0.01$. See text for further details.
   For the NH and QD (and interpolating) spectra, the regions within
   the black lines of a given type (solid, short-dashed, long-dashed,
   dash-dotted) correspond to the four different sets of CP-conserving
   values of the two phases $\alpha_{21}$ and $\alpha_{31}$, and thus
   to the four possible combinations of the relative CP parities
   ($\eta_{21},\eta_{31}$) of neutrinos $\nu_{1,2}$ and $\nu_{1,3}$:
   $(+1,+1)$ solid, $(-1,-1)$ short-dashed, $(+1,-1)$ long-dashed, and
   $(-1,+1)$ dash-dotted lines.  For the IH spectrum, the regions
   delimited by the black solid (dotted) lines correspond to
   $\eta_{21} = + 1$ ($\eta_{21} = -1$), independently of $\eta_{31}$.
   The regions shown in red correspond to violation of
   CP-symmetry.}
\end{figure}

   In Fig.~\ref{fig:meffbf} we show 
the predicted ranges of $\meff$ 
using the present best fit values of the neutrino 
oscillation parameters and their prospective errors
as discussed above. We assumed a 1$\sigma$ experimental error 
of 2\%, 2\% and 4\% on  $\deltasol \!\!$, $\deltaatm$, 
$\sin^2 \theta_\odot$, respectively. 
For $\sin^2 \theta_{13}$, we take $\sin^2 \theta_{13}=0.01$ and we consider 
the 1$\sigma$ uncertainty in the absolute value 
of 0.006.
%
In Table 2 we give the maximal and minimal 
values of $\meff$ for the three spectra, 
NH, IH and QD.
\begin{table}[ht]
\begin{center}
\begin{tabular}{|c|c|c|c|c|} 
\hline
\rule{0pt}{0.5cm} 
$\meff_{\rm min}^{\rm NH}$ & $\meff_{\rm max}^{\rm NH}$ & $\meff_{\rm min}^{\rm IH}$ &  $\meff_{\rm max}^{\rm IH}$ &
$\meff_{\rm min}^{\rm QD} $  
\\ \hline \hline
2.1 \! [1.5] \! (1.0) & 3.5 \! [3.9 ] \! (4.4)  &  15.1 \! [15.0] \! (14.8)    &  50.1 \! [50.0] \! (49.1)   & 63.4 \! [60.7] \! (58.0) \\ \hline
\end{tabular}
\caption{\label{tab:bf} 
The maximal values of  
\meff{} (in units of meV)
for the NH  and IH spectra, 
and the minimal values of 
\meff{} (in units of meV) 
for the NH, IH and QD spectra,
for the best fit values 
of the oscillation parameters
and using the prospective errors 
discussed in the text.
We take $\sin^2 \theta_{13} = 0.0 \ [0.01] \ (0.02)$.
The results for the NH and IH spectra 
are obtained for
$\mmin = 10^{-4}$ eV, while
those for the QD spectrum correspond to 
$\mmin = 0.2$ eV\@.
}
\end{center}
\end{table}

\section{The $\meff \sim 10^{-3}$ eV Frontier in $\betabeta-$Decay}


 In the present Section we will analise the conditions 
under which $\meff \gtap 10^{-3}$ eV in the case 
of neutrino mass spectrum with normal ordering.
Before discussing the general case 
of arbitray $m_1$ and $\sin^2\theta_{13}$ 
satisfying the presently exiting experimental 
limits, we will consider two 
specific but physically interesting cases: 
i) negligibly small $m_1$ (NH spectrum), 
and ii) relatively small $\sin^2\theta_{13}$, 
such that the term     
$\sqrt{m_1^2 + \deltaatm} \sin^2\theta_{13}$ 
in Eq. (\ref{meffcompleteNH}) is strongly suppressed, 
$\sqrt{m_1^2 + \deltaatm} \sin^2\theta_{13} \ltap 10^{-4}$ eV. 

%
\subsection{Normal Hierarchical Spectrum}
%
%

  In the case of the normal hierarchical spectrum 
we have $m_1 \ll m_{2,3}$ and therefore 
only the two heavier neutrinos, $\nu_2$ and $\nu_3$,
contribute to the effective Majorana mass parameter.
In this case $m_2 \cong \sqrt{\deltasol}$, 
$m_3 \cong \sqrt{\deltaatm}$, and
the sum of neutrino masses reads: 
\beq
m_1 + m_{2} +  m_{3} \cong 0.058~{\rm eV}\,.
\label{NHsum}
\eeq
%
\noindent The effective Majorana mass is given by:
\begin{equation}
\label{meffNH1}
 \meff \simeq \left| \sqrt{\deltasol} 
\sin^2 \theta_\odot (1- \sin^2 \theta_{13})
 + \sqrt{\deltaatm} \sin^2 \theta_{13} e^{i \alpha_{32}} \right|\,,
 \end{equation}
%
where $\alpha_{32}\equiv \alpha_{31} - \alpha_{21}$ 
is the difference of the two Majorana 
CP-violating phases in $\pmns$. 
We will refer to the first term in the 
r.h.s. of Eq.~(\ref{meffNH1}) as the ``solar term''
due to its dependence on $\deltasol$, 
while to the second as the ``atmospheric'' one. 
The two terms in the expression for $\meff$ 
add constructively if $0 \leq \alpha_{32} \leq \pi/2$, 
while for $\pi/2 < \alpha_{32} \leq \pi$ 
partial or complete cancellation 
between the ``solar'' and ``atmospheric'' 
terms can take place.
The cancellation is most effective 
in the case of CP-invariance and
$\alpha_{32} = \pi$.
The degree of cancellation is controlled 
by $\sin^2 \theta_{13}$.
For sufficiently small values of $\theta_{13}$,
$\sin^2\theta_{13} \ltap 0.01$, 
the solar term dominates
and $\meff$ is predicted to be in the few meV range, 
$\meff \sim (2 - 3)\times 10^{-3}$ eV. 
If $\sin^2 \theta_{13}$ is close 
to the present 3$\sigma$ bound~
\cite{TSchwNuFact07}, $\sin^2 \theta_{13} < 0.05$,
the solar and the atmospheric terms in Eq.~(\ref{meffNH1})
are of the same order and a substantial cancellation 
can take place. We will analise this possibility first 
qualitatively. 

   Consider the ``extreme'' case 
of $\alpha_{32}=\pi$ and $\meff = 0$~ 
\footnote{We postpone the discussion 
of the $\betabeta-$decay 
in the case of $\meff = 0$ to Section 4.}. 
This requires \cite{BPP1,PPW,LMWR05}
\beq
\meff = 0:~~~~~~~~~
\sin^2\theta_{13} = \frac{\sqrt{\deltasol}}{\sqrt{\deltaatm}}\, 
\sin^2 \theta_\odot\,,~~~~~~~
\label{NHmeff0}
\eeq
%
\noindent
where we have neglected \deltasol
with respect to \deltaatm.
Taking the best fit values of $\deltasol$, 
$\sin^2 \theta_\odot$ and $\deltaatm$, 
determined from the analysis of the 
currently existing neutrino oscillation 
data, we get $\sin^2\theta_{13} = 0.057$, 
which is ruled out by the data.
Using the 2$\sigma$ and 3$\sigma$ 
ranges of allowed values of 
the same three parameters, we find 
respectively $\sin^2\theta_{13} = 0.046$,
which is close to the current 3$\sigma$ 
upper limit on $\sin^2\theta_{13}$,
and $\sin^2\theta_{13} = 0.041$.
Thus, in order for $\meff$ to be strongly 
suppressed, $\meff \ll 10^{-3}$ eV, 
$\sin^2\theta_{13}$ should have a value 
close to the existing 3$\sigma$ 
upper limit. If we use the current  
2$\sigma$  (3$\sigma$) upper limit on 
$\sin^2\theta_{13}$, 
$\sin^2\theta_{13} < 0.033~(0.050)$, 
and the present best fit values 
of  $\deltasol$, $\sin^2 \theta_\odot$ 
and $\deltaatm$, we find for $\alpha_{32} = \pi$
that $\meff \gtap 1.1~(0.2)\times 10^{-3}$ eV.
If  $0 \leq \alpha_{32} \leq 5\pi/6$, we obtain
$\meff \gtap 1.5~(1.3)\times 10^{-3}$ eV.
It follows from this simple analysis that 
if, in the future high precision measurements 
of $\deltasol\!\!$, $\sin^2 \theta_\odot$ 
and $\deltaatm$, the currently determined 
best fit values of these parameters will not 
change and $\sin^2\theta_{13}$ is found to 
have a value $\sin^2\theta_{13} \ltap 0.01~(0.03)$,
the effective Majorana mass will 
satisfy $\meff \gtap 2.2~(1.2)\times 10^{-3}$~eV 
for any $\alpha_{32}$.
For, e.g. $0 \leq \alpha_{32} \leq 5\pi/6$,
we have $\meff \gtap 1.3\times 10^{-3}$ eV 
for any $\sin^2\theta_{13}$
allowed at 3$\sigma$ by the existing data.
Values of $\alpha_{32}\neq 0$  in the indicated 
range are required for the generation 
of the baryon asymmetry of the Universe 
in the ``flavoured'' leptogenesis scenario, 
in which the requisite CP-violation is provided 
exclusively by the Majorana phase (difference) 
 $\alpha_{32}$ \cite{PPRio106}.

  We will perform next a similar 
analysis of the conditions 
under which $\meff \gtap 10^{-3}$ eV, taking 
into account the current and prospective 
uncertainties in the measured values 
of the relevant neutrino oscillation 
parameters. The minimal predicted value of 
\meff, \meffmin, is obtained in the case of 
CP-conservation and opposite CP-parities 
of the two relevant neutrinos and 
can be evaluated as
\begin{equation}
\meffmin = \meff_-    -  n \sigma(\meff_-),
\end{equation}
%
where $\meff_-$ is the predicted value of 
\meff obtained using the best fit values of the
oscillation parameters, 
$\sigma(\meff_-)$ is the error on \meff and $n=1,2,3...$.
 
  Using the propagation of errors 
and assuming that the errors on the 
oscillation parameters of interest 
are small and independent,
we obtain the 1-$\sigma$ error on $\meff$
for any $\alpha_{32}$:
\begin{equation}
 \label{sigmaalpha}
 \begin{tabular}{ll}
$ \sigma(\meff \! \!) \simeq $ & \hspace{-0.5truecm}
$ {\displaystyle\frac{1}{2 \meff\! \! }  
\left(  \sin^4 \theta_{13} \, \deltaatm
\Big( \sin^2 \theta_{13} \sqrt{\deltaatm \! }  
+ \sqrt{\deltasol \! } \sin^2 \theta_\odot \cos \alpha_{32}  \Big)^2
   \delta^2(\deltaatm) \right.}$ \\ [0.25cm]
  & \hspace{-2.5truecm}  
$ + \, \deltasol  \sin^4 \theta_\odot   
 \left( \sqrt{\deltasol\!} \sin^2 \theta_\odot  +  
\sqrt{\deltaatm} \sin^2 \theta_{13} \cos \alpha_{32}  \right)^2
 \Big( 4 \delta^2(\sin^2 \theta_\odot) + \delta^2(\deltasol\!) \Big)$\\[0.25cm]
  & \hspace{-2.5truecm} $ \left. + 
4\left( \sin^2 \theta_{13} \deltaatm \! \! 
- \deltasol \! \! \sin^4 \theta_\odot 
+ \sqrt{\deltasol \! \! \deltaatm} 
\sin^2 \theta_\odot \cos \alpha_{32} \right)^2 \! \!  
\sigma^2(\sin^2 \theta_{13}) \! \! 
\right)^{1/2} $\,.
\end{tabular}
\end{equation}
%
Here $\delta(\sin^2 \theta_\odot), 
\delta(\deltasol\!\!)$ and $\delta(\deltaatm)$
are the relative errors on 
the oscillation parameters $\dmsol$, $\sin^2\theta_{\odot}$
and $\deltaatm$, $\sigma(\sin^2 \theta_{13})$ is the absolute 
error on $\sin^2 \theta_{13}$, and
we have used the fact that $\sin^2 \theta_{13} \ll 1$.
We have assumed (see Section~2 and Fig.~2) 
and will use in our further 
analysis (see Section 2) 
the following values of 
the errors: 
$\delta(\sin^2 \theta_\odot) = 4\%$, 
$\delta(\deltasol\!\!) = 2\%$ and 
$\delta(\deltaatm) = 2\%$.
For the chosen values 
$\delta(\sin^2 \theta_\odot), 
\delta(\deltasol\!\!)$ and $\delta(\deltaatm)$,
the error on  $\deltasol$ 
gives a subdominant contribution 
in comparison with that on
the solar mixing angle and we neglect 
it in the following discussion.

  If CP-invariance holds we have
$\alpha_{32} = 0, \pi$ and 
Eq.~(\ref{sigmaalpha}) simplifies to:
\begin{equation}
\label{sigmaCP}
\sigma(\meff_{\pm}) \simeq 
\sqrt{ \deltasol \sin^4 \theta_\odot   \delta^2(\sin^2 \theta_\odot)  
+ \displaystyle\frac{\sin^4 \theta_{13} \deltaatm}{4} 
\delta^2(\deltaatm) + \deltaatm \sigma^2(\sin^2 \theta_{13}) }\,,
\end{equation}
%
where we have neglected 
$\sqrt{\deltasol} \sin^2 \theta_\odot$ with respect to
$\sqrt{\deltaatm}$. In Eq. (\ref{sigmaCP})
$\meff_{\pm}$ refers to $\eta_{32}=\pm 1$. 
The contribution of the error on $\deltaatm$ 
in $\sigma(\meff_{\pm})$ 
is suppressed by the factor 
$\sin^2 \theta_{13}$ and 
can also be neglected, while the errors on 
$\sin^2 \theta_{13}$  and on 
$\sin^2 \theta_\odot$ can give sizable 
contributions to $ \sigma(\meff_{\pm})$
and both should be taken into account. 
For the current best 
fit values of the oscillation parameters,
$\sigma(\meff_{\pm})$ is given 
to a good approximation by
$ \sigma(\meff_{\pm}) \cong \sqrt{\deltaatm} 
\sqrt{ (0.057 \delta^2(\sin^2 \theta_\odot)) +  
\sigma^2(\sin^2 \theta_{13})}$. 
It is clear from this expression that 
for an error on $\sin^2 \theta_\odot$ of 4--8\%,
the two terms in $\sigma(\meff_{\pm})$ 
are of the same order 
if $\sigma(\sin^2 \theta_{13}) = 0.004$, while
for $\sigma(\sin^2 \theta_{13})\gtap 0.006$ 
the error on $\sin^2\theta_{13}$ 
typically gives the dominant 
contribution in $\sigma(\meff_{\pm})$.

  For neutrinos of equal CP-parities, i.e. $\alpha_{32} = 0$,
the mean value of $\meff$ is predicted to be 
in the few meV range and 
the expected relative error $\sigma(\meff_{\pm})$
varies between 7\% and 15\%, depending on the specific values 
of errors and best fit values of the 
parameters. If 
the neutrinos $\nu_2$ and $\nu_3$
have opposite CP-parities, i.e. $\alpha_{32} = \pi$,
the mean value of \meff is smaller as partial 
cancellation between their contributions to \meff 
can take place. In this case the error on \meff
can become as large as 30\%--40\%.

  If CP-symmetry is broken, 
the full expression for $\sigma(\meff)$, 
Eq.~(\ref{sigmaalpha}), should be used.
It can be shown, however, that
$\sigma(\meff) < \mathrm{max} 
\Big(  \sigma(\meff_{+}),  \sigma(\meff_{-}) \Big)$.

   Using Eq.~(\ref{sigmaCP}) in the case of $\eta_{32} = -1$,
we can study analytically the condition
on $\sin^2 \theta_{13}$ which guarantees that
the predicted value of \meff is larger than 1~meV. 
Neglecting the dependence on $\sin^2 \theta_{13}$ in 
$\sigma(\meff)$, we find an approximate 
solution for $\sin^2 \theta_{13}$:
\begin{equation}
\label{xapprox}
\sin^2 \theta_{13} < \displaystyle\frac{\sqrt{\deltasol} \sin^2 \theta_\odot  - 1 \ {\rm meV} - n \sqrt{\deltasol \sin^4 \theta_\odot \delta^2(\sin^2 \theta_\odot) + \deltaatm \sigma^2(\sin^2 \theta_{13})}}{\sqrt{\deltaatm}}.
\end{equation}

 \begin{figure}
 \label{fig:sinthetabf}
\centerline{\begin{tabular}{c}
\includegraphics*[height=0.4\textheight, width=0.8\textwidth]{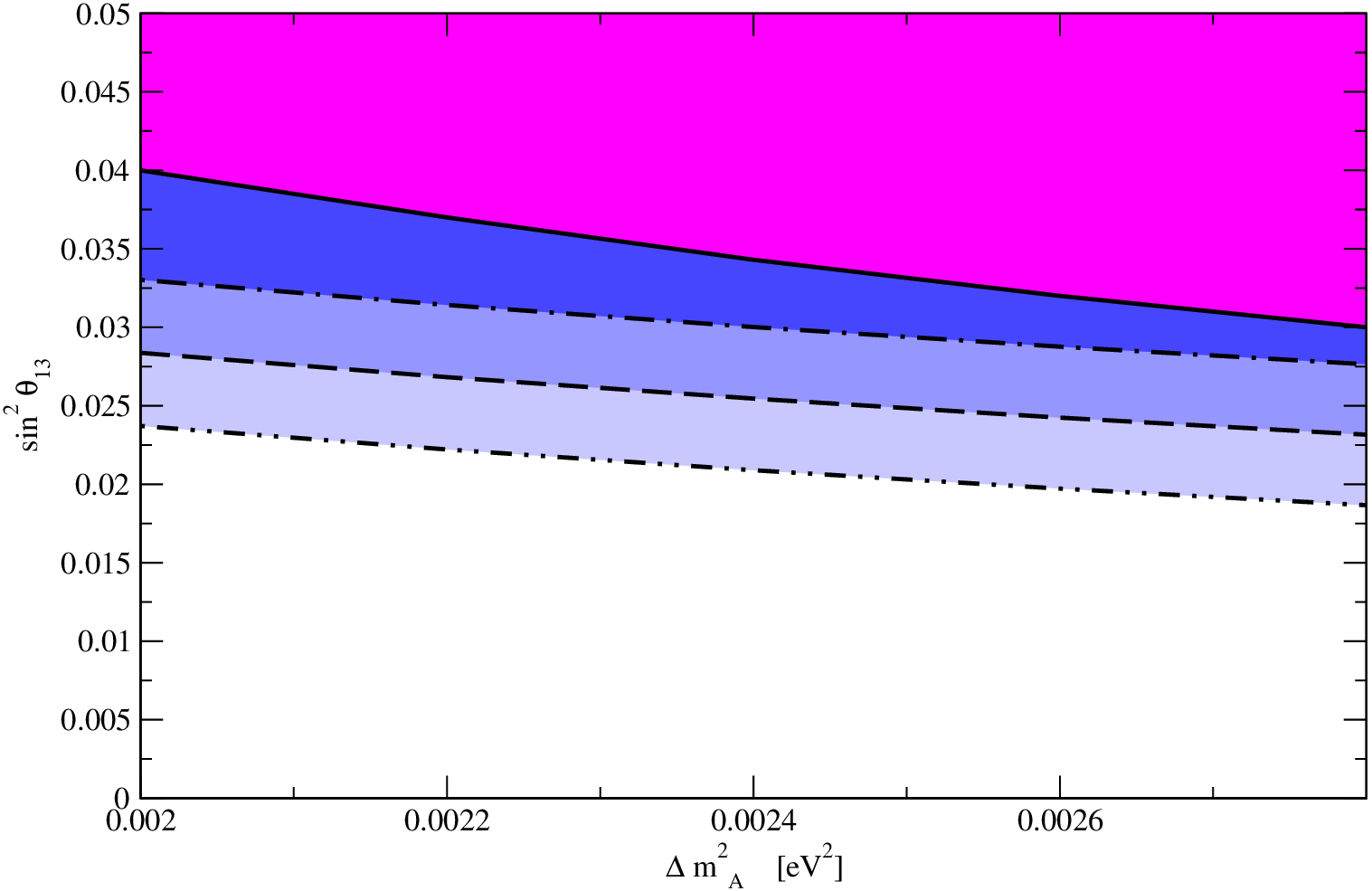} \\ 
\includegraphics*[height=0.4\textheight, width=0.8\textwidth]{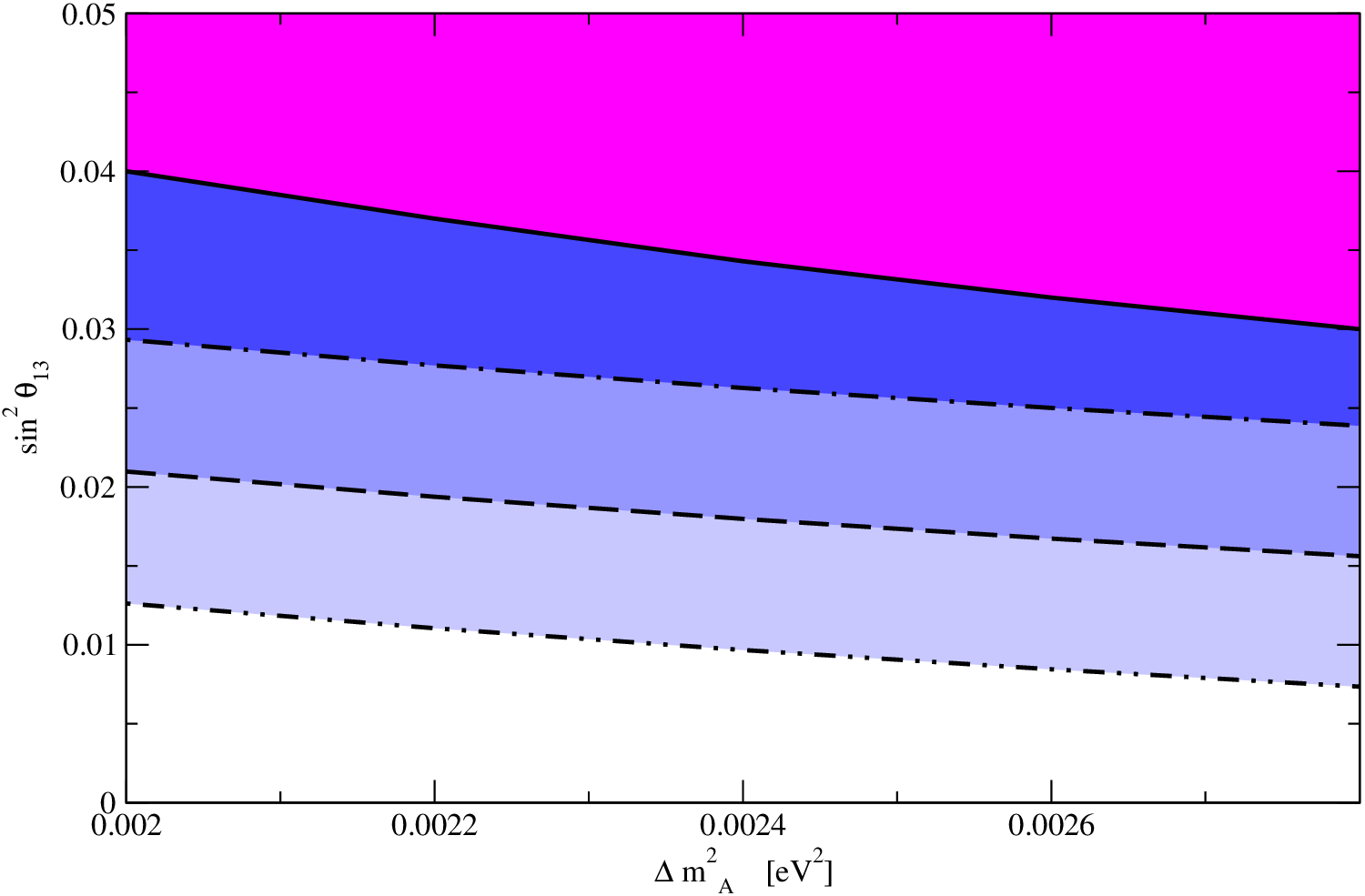} 
\end{tabular}}
\caption{ The coloured regions show the values of
$\sin^2 \theta_{13}$ versus $\deltaatm$ for which
$\meffnhmin <1~$meV at 1 (2) [3] $\sigma$ 
(region bounded from below by the dash-dotted
(dashed) [dash-double-dotted] line for
$\sin^2 \theta_\odot = 0.32$.
The error on $\sin^2 \theta_{13}$ is taken to be 0.004
(0.008) in the upper (lower) plot.
The medium-grey (magenta) region is excluded by the present bound
on $\sin^2 \theta_{13}$~
\protect\cite{TSchwNuFact07}.
}
\end{figure}
%
 In Fig.~3 we show the values of 
$\sin^2 \theta_{13}$ versus \deltaatm~ for which 
$\meffmin = 1~$meV is satisfied
for $n$=1, 2, 3 (dash-dotted, dashed, dash-double-dotted lines).
We use the best fit value of $\sin^2 \theta_\odot$
and two values of the error on $\sin^2 \theta_{13}$. 
 If $\sin^2 \theta_{13}$ is larger than 
 the shown values, a strong cancellation between
 the two contributions to \meff can take place and 
 $\meffmin < 1~$meV. This would imply that,
 depending  on the value of $\alpha_{32}$,
 there are predicted values of \meff 
 both smaller and larger than the future reference sensitivity 
used in this analysis. The possibility for a future experiment 
to find a positive signal of \betabeta-decay would depend 
 on the unknown value of $\alpha_{32}$.
 
   The limiting value of $\sin^2 \theta_{13}$ 
is in the 0.01--0.03 range.
The precise value depends critically on the error
on $\sin^2 \theta_{13}$:
for $\sigma(\sin^2 \theta_{13}) \simeq 0.004 \ (0.008)$, 
we have $\sin^2 \theta_{13}  < 0.02 \ (0.01)$.
The limit on $\sin^2 \theta_{13}$
depends also on \deltaatm, as can be easily 
understood from Eq.~(\ref{xapprox}):
the larger $\deltaatm$, the smaller the bound on
$\sin^2 \theta_{13}$.
The value of $\sin^2 \theta_\odot$
controls the magnitude of the
first term in $\meff_-$ and therefore 
plays an important role in Eq.~(\ref{xapprox}).
We show the dependence on  $\sin^2 \theta_\odot$ in 
Figs.~4 and 5.
 \begin{figure}
 \label{fig:sinthetasol2}
\centerline{\begin{tabular}{c}
\includegraphics*[height=0.46 \textheight, width=0.75 \textwidth]{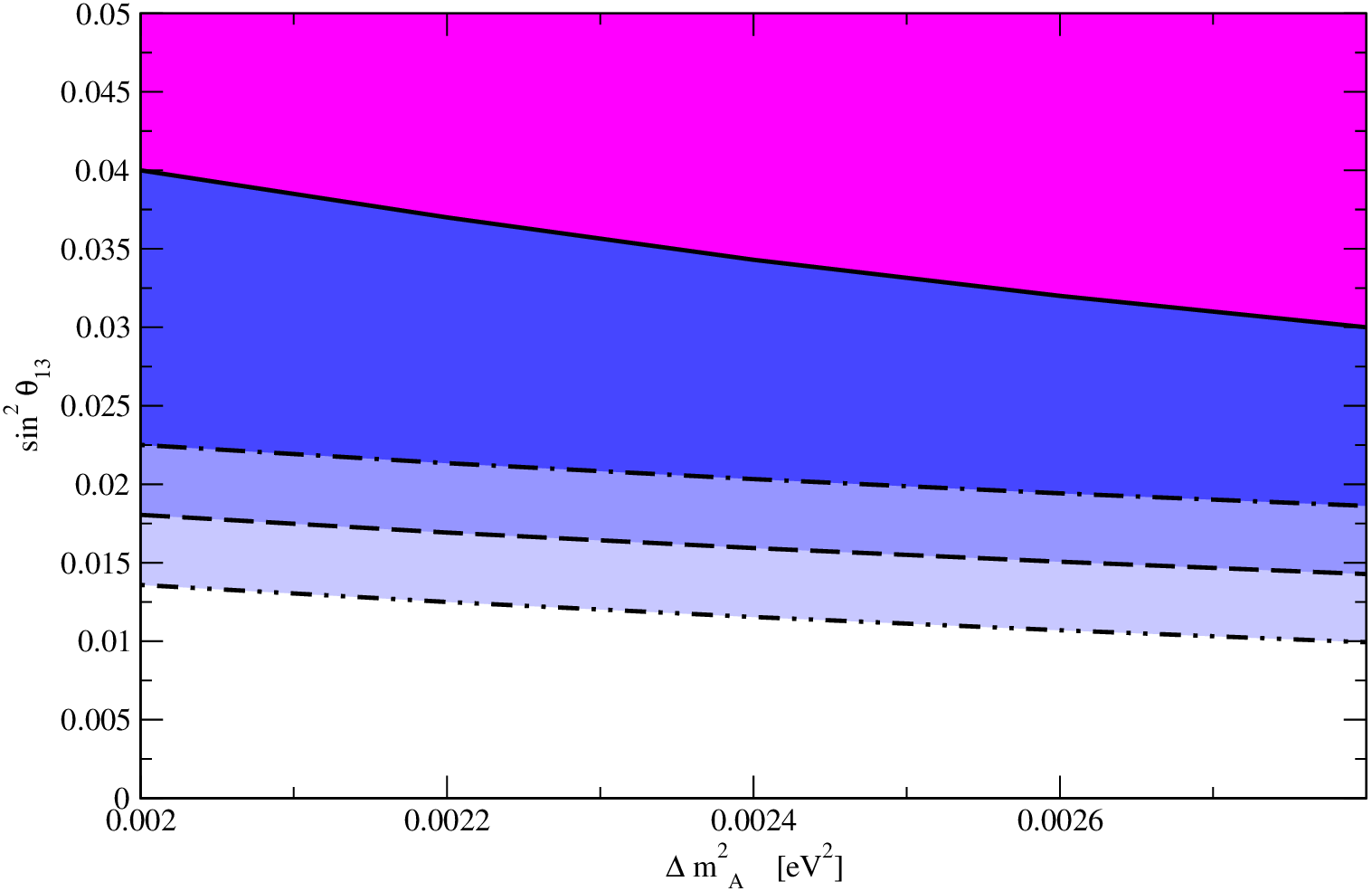} \\
\includegraphics*[height=0.46 \textheight, width=0.75 \textwidth]{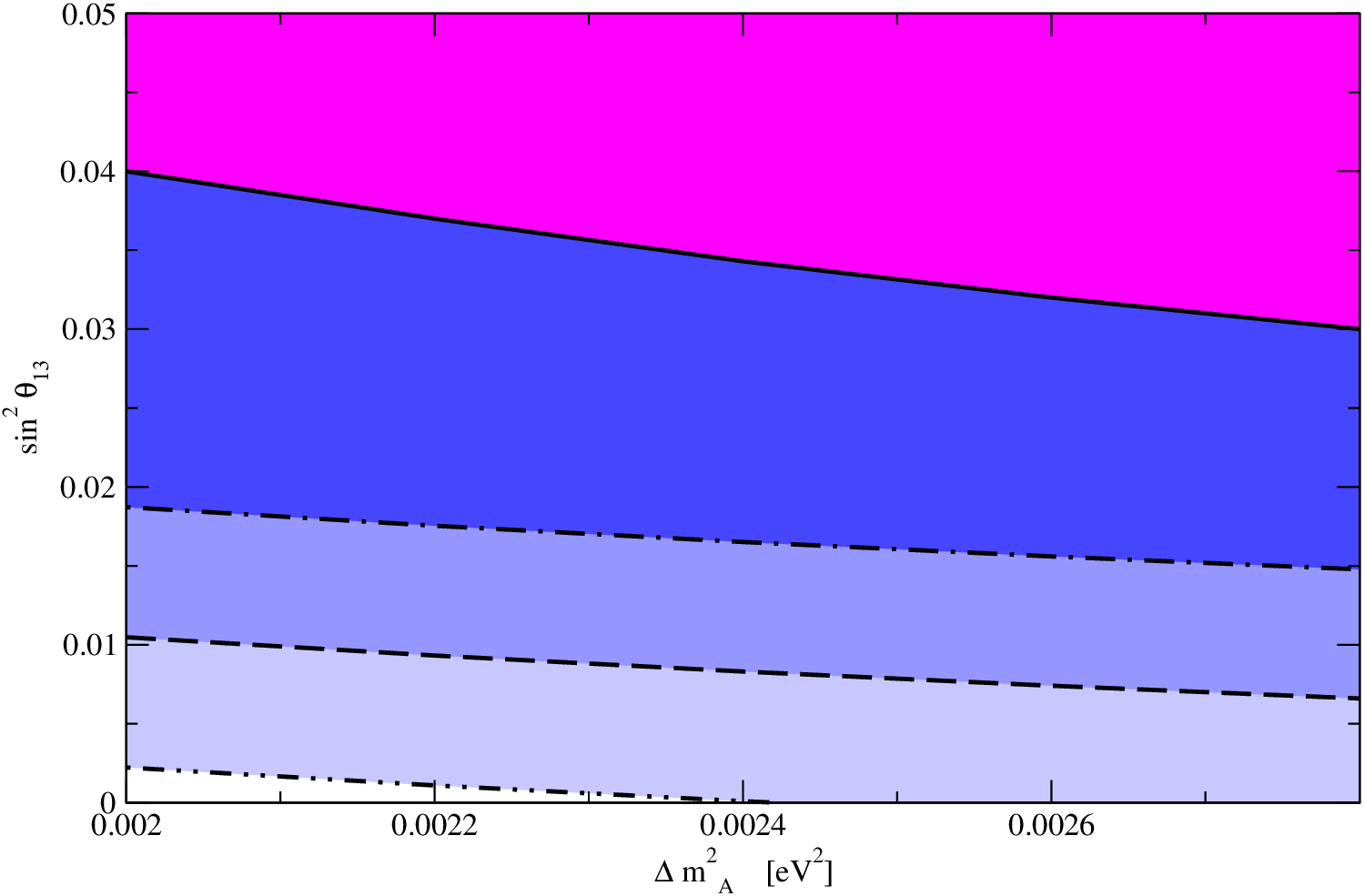} 
\end{tabular}}
\caption{The same as in Fig.~3,
but for $\sin^2 \theta_\odot = 0.26$.
}
\end{figure}
 \begin{figure}
 \label{fig:sinthetasol3}
\centerline{\begin{tabular}{c}
\includegraphics*[height=0.46 \textheight, width=0.75 \textwidth]{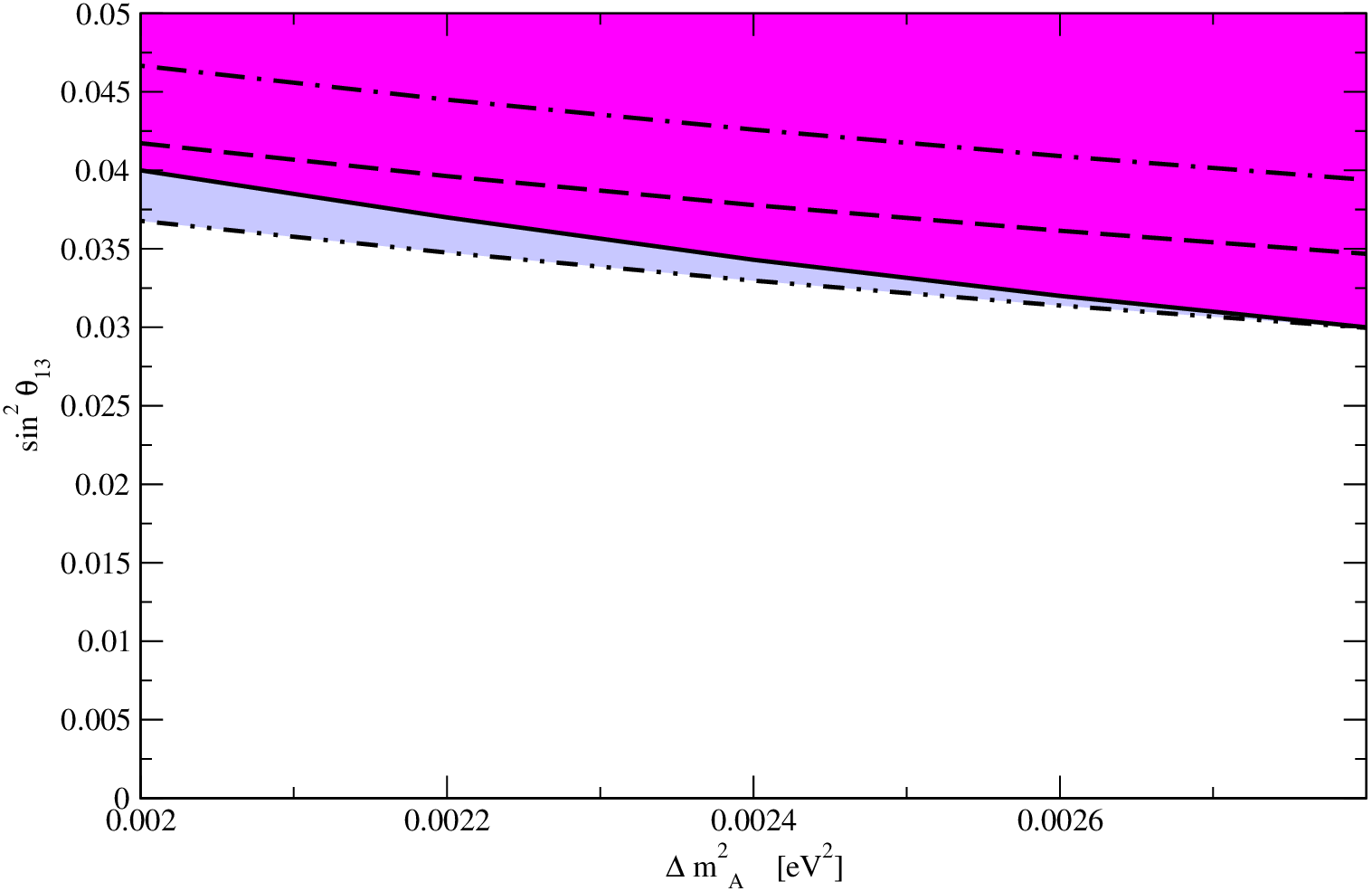} \\
\includegraphics*[height=0.46 \textheight, width=0.75 \textwidth]{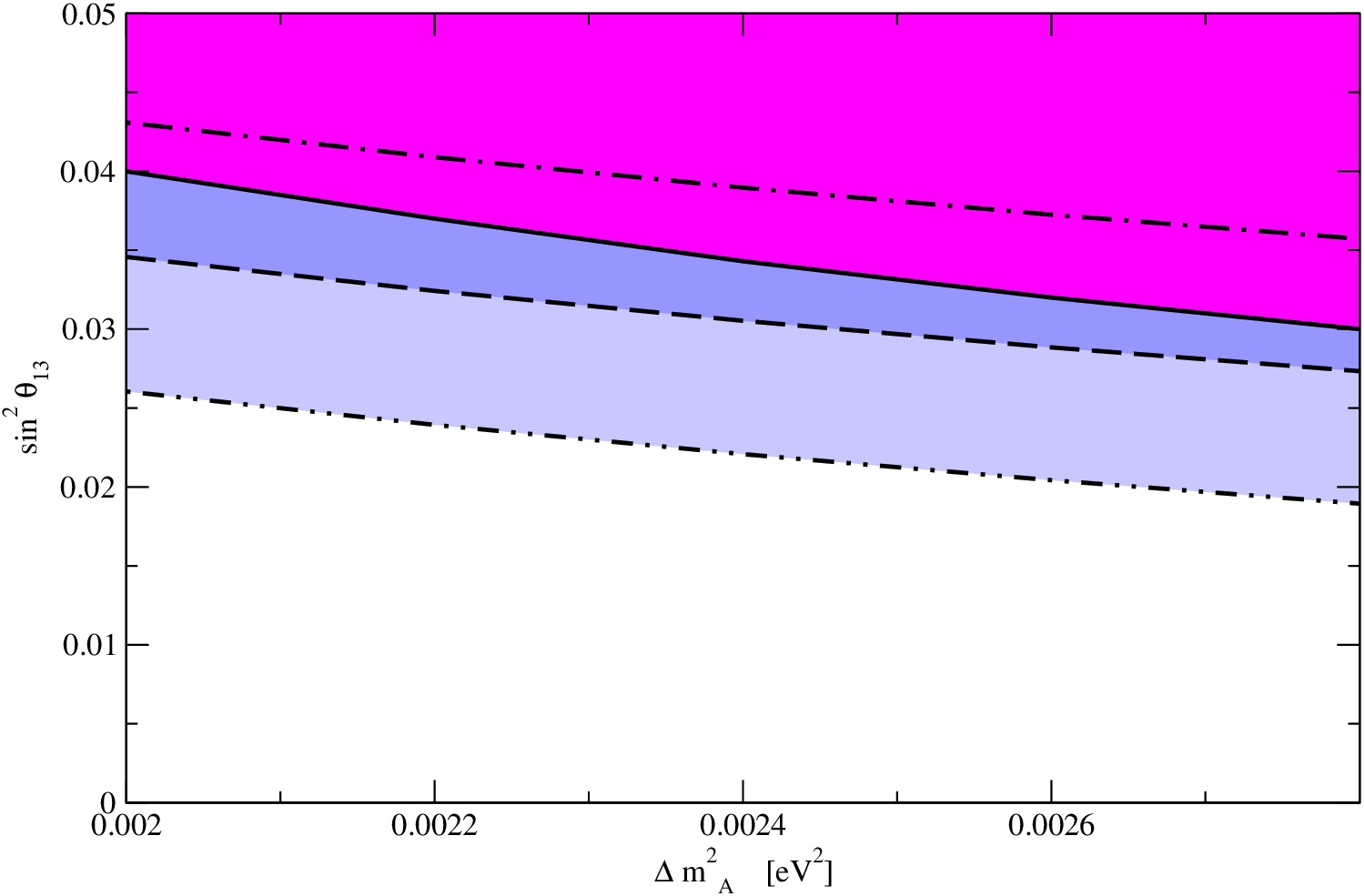}
\end{tabular}}
\caption{The same as in Fig.~3,
but for $\sin^2 \theta_\odot = 0.40$.
}
\end{figure}
%
The smaller the value of $\sin^2 \theta_\odot$,
the smaller $\sin^2 \theta_{13}$
for which one can have $\meffmin < 1$~meV.
If $\sin^2 \theta_\odot = 0.26$ and 
$\sigma(\sin^2 \theta_{13}) \simeq 0.004$,
we have $\meff > 0.001$ eV
for values of $\sin^2 \theta_{13} < 0.01$. 
If, however, $\sigma(\sin^2 \theta_{13}) \simeq 0.008$,
one can have $\meffmin < 1$~meV even 
if the (mean)  value of 
$\sin^2 \theta_{13} = 0$.
On the contrary, for $\sin^2 \theta_\odot = 0.40$,
a large part of the relevant 
parameter space is already
excluded by the present data~
\cite{TSchwNuFact07}
and we get $\meff > 0.001$ eV for
$\sin^2 \theta_{13} < 0.03 \ (0.02)$ 
in the case of 
$\sigma(\sin^2 \theta_{13})=0.004 \ (0.008)$.

 The preceding rather detailed analysis shows 
that  $\meff \geq 0.001$ eV  typically
for $\sin^2 \theta_{13} \ltap (0.01 - 0.02)$.
Values of $\sin^2 \theta_{13} \gtap (0.01 - 0.02)$ 
are within the sensitivity of the two reactor experiments 
Double-CHOOZ~\cite{DCHOOZ} and Daya Bay \cite{DayaB},
which are under preparation, and of
the currently operating and future
long baseline neutrino oscillation experiments,
MINOS~\cite{MINOS}, OPERA~\cite{Opera},
T2K and NO$\nu$A \cite{Future}.
The results of these experiments will be crucial 
for establishing whether the effective Majorana mass 
$\meff$ in the case of NH neutrino mass spectrum
is limited from below and for determining its lower 
limit.  

   In the case of a NH spectrum,
for $\sin^2 \theta_{13} = 0$, 
only one contribution in $\meff$ is relevant, 
the other two being suppressed 
by the negligible values of $m_1$ and $\sin^2 \theta_{13}$.
In this case there is no dependence of \meff on $\alpha_{32}$.
If $\sin^2 \theta_{13}$ has a value close 
to the existing upper limit,
a sufficiently accurate measurement of \meff 
could allow to distinguish the two possible CP-parity 
patterns or establish CP-violation.
Here, we study what would be the requirements in order to have
sensitivity to CP-violation.
We perform a simplified analysis in which 
we retain for both CP-parity patters
only the dominant term in the theoretical 
error on $\meff$:
\begin{equation}
\sigma(\meff) \simeq \sqrt{\deltaatm} \sigma(\sin^2 \theta_{13})\,.
\end{equation}
%
The existence of a "just-CP-violating" region~\cite{BPP1},
signaling the possibility to search for CP-violation, 
requires the allowed regions for the CP-conserving
cases $\eta_{32} = 1$ and $\eta_{32} = -1$
not to overlap. This condition is satisfied provided
\begin{equation}
\label{s13condition}
\sin^2 \theta_{13} > n \sigma(\sin^2 \theta_{13}),
\end{equation}
%
where $n$ is the number of $\sigma(\sin^2 \theta_{13})$ considered.
For example, for $\sigma(\sin^2 \theta_{13}) = 0.004, 0.008$
and $n=2$, we have $\sin^2 \theta_{13} > 0.008, 0.016$.
In this case, in principle, it would 
be possible to distinguish the two CP-parities 
patterns or find CP-violation due to a Majorana 
CP-violating phase.
CP-violation would be established 
if the experimentally allowed value
of \meff is within the "just-CP-violating" region,
once the experimental error on \meffnospace, $\Delta$, 
and the nuclear matrix elements 
uncertainties are taken into account.
Even if Eq.~(\ref{s13condition}) is satisfied,
this is a formidably challenging task.
In the most optimistic case of $\sin^2 \theta_{13}$
having a value close to the present 3$\sigma$ bound, 
$ \sin^2 \theta_{13} \simeq 0.05$,
for a nuclear matrix element uncertainty
$\zeta = 1.5$ on \meffnospace,
an error not larger than $\Delta = 0.5$~meV 
would be required.
The width of the "just-CP-violating" region
decreases rapidly with $\theta_{13}$ and  
for smaller values of $\sin^2 \theta_{13}$
the error required on \meff would be even smaller.


\subsection{The Case of Small $\sin^2\theta_{13}$}


 Consider next the possibility of $\sin^2\theta_{13}$ 
having a rather small value, such that\\ 
\noindent $\sqrt{m_1^2 + \deltaatm} \sin^2\theta_{13} 
\leq 2\times 10^{-4}~{\rm eV} \ll 10^{-3}$ eV.
For $m_1^2 \ll \deltaatm$ this conditions is fulfilled 
if $\sin^2\theta_{13} \ltap 4\times 10^{-3}$, 
while if, e.g. $m_1 \cong 0.05$ eV, it is satisfied 
provided $\sin^2\theta_{13} \ltap 3\times 10^{-3}$.
These values of $\sin^2\theta_{13}$ can be tested, e.g.
in future long baseline neutrino experiments
with superbeams, beta beams and
at neutrino factories~\cite{reactoaccfuture,Blondel:2006su}.

 We set $\sin^2\theta_{13} = 0$ for simplicity 
in the following discussion. The expression for 
$\meff$ simplifies to: 
\begin{equation}
\meff = \left| m_1\, \cos^2 \theta_\odot 
+ \sqrt{m_1^2 + \deltasol}\, \sin^2 \theta_\odot\, 
e^{i \alpha_{21}} \right |\,.
\label{meff1NHth130}
\end{equation}
%
For NH neutrino mass spectrum, i.e. for 
$m_1 \ll \sqrt{\deltasol} 
\sin^2 \theta_\odot/\cos^2 \theta_\odot \cong 4\times 10^{-3}$ eV,
we always have $\meff \cong 3\times 10^{-3}$ eV.
If, however, $m^2_1 \gtap \deltasol$, the neutrino 
mass spectrum will not be hierarchical. 
There are two possibilities.

  {\bf i)} For $m^2_1 \gg \deltasol \cong 7.6 \times 10^{-5}~{\rm eV^2}$,
we get 
\begin{equation}
\meff \cong m_1\,\sqrt{ 1 - \sin^22\theta_{\odot}\, 
 \sin^2\frac{\alpha_{21}}{2} } \gtap 
 m_1\, \cos2\theta_{\odot}\,.
\label{meff2NHth130}
\end{equation}
%
Taking $m_1 \gtap 2\times 10^{-2}$ eV and the 
2$\sigma$ (3$\sigma$) lower limit on $\cos2\theta_{\odot}$,
$\cos2\theta_{\odot} \geq 0.26~(0.20)$, we find 
$\meff \gtap 5.2~(4.0)\times 10^{-3}$ eV.
In this case $m_2 = \sqrt{m_1^2 + \deltasol} 
\gtap 2.2\times 10^{-2}$ eV, and the sum of neutrino masses 
satisfies:
\begin{equation}
m_1 + m_2 + m_3 \gtap 9.5\times 10^{-2}~ \mathrm{eV}\,.
\label{sum1NHth130}
\end{equation}
%
{\bf ii)} If, however, $m^2_1 \sim \deltasol$ and $\alpha_{21} \sim \pi$,
a cancellation between the two terms in Eq. (\ref{meff1NHth130})
is possible and $\meff$ can be strongly suppressed,
$\meff \ll 10^{-3}$ eV. 
Consider the extreme case of $\meff = 0$ (for a more detailed 
discussion of the $\betabeta-$decay in the case of $\meff = 0$ see 
Section 4). For $\alpha_{21} = \pi$, it is realised if~\cite{PPW,LMWR05}
\begin{equation}
\meff = 0:~~~~~~m_1 = m_2\, \tan^2\theta_{\odot}\,.~~~~~~~~~~
\label{NHth130meff0}
\end{equation}
%
Using the relation $m_2 = (m_1^2 + \deltasol)^{\frac{1}{2}}$, 
we find that $\meff = 0$ can hold in the case being studied
if $m^2_1 = \deltasol \sin^4 \theta_\odot/\cos2 \theta_\odot 
\cong  2.2\times 10^{-5}~{\rm eV^2}$, where we have used the 
best fit values of $\deltasol$ and $\sin^2 \theta_\odot$.
This implies that $m_1 \cong 4.6\times 10^{-3}$ eV, 
$m_2 \cong 10^{-2}$ eV, and, correspondingly, 
\begin{equation}
m_1 + m_2 + m_3 \cong 6.4\times 10^{-2}~ \mathrm{eV}\,.
\label{sum2NHth130}
\end{equation}
%
It is not difficult to convince oneself, however, 
that if $\alpha_{21} = \pi$,
one obtains $\meff \gtap \mu $ for 
\begin{equation}
m_1\gtap \frac{\mu}{\cos2\theta_{\odot}}\,
\left [ \cos^2\theta_{\odot} + \sin^2\theta_{\odot}\, 
\sqrt{1 + \mu^{-2}\, \deltasol\,\cos2\theta_{\odot}} \right ]\,,~~
\label{m1a}
\end{equation}
%
where the reference value 
$\mu = 10^{-3}~{\rm eV}$ 
in the case of interest. 
Using the best fit values of 
$\deltasol$ and $\sin^2\theta_{\odot}$
we get $m_1 \gtap 6.6\times 10^{-3}~{\rm eV}$.
For the sum of neutrino masses we obtain
$m_1 + m_2 + m_3 \gtap 6.7\times 10^{-2}$ eV.

  This qualitative analysis 
shows that if $\sin^2\theta_{13} \ltap 3\times 10^{-3}$
and the sum of neutrino masses satisfies 
$m_1 + m_2 + m_3 \gtap 7\times 10^{-2}$ eV,
we will have $\meff \geq 10^{-3}$ eV for any 
$\alpha_{21}$.


\subsection{Spectrum with Partial Hierarchy }


  As is well-known \cite{PPW}, 
in the case of neutrino mass spectrum with partial 
hierarchy we can have $\meff \ll 1$~meV
and even $\meff = 0$. However, 
this requires that the lightest 
neutrino mass $m_1$ has a value 
in the rather narrow interval,
$m_1\sim ({\rm few}\times 10^{-3} - 10^{-2}$) eV.
As a consequence, the sum 
of neutrino masses should 
also lie within a specific interval.
Here we analyze the values of
$m_1$ and $\sin^2\theta_{13}$ 
for which the indicated strong cancellation
in $\meff$ would not take place and we would have
$\meff \geq 1$~meV.

  For the neutrino mass spectrum under discussion,
all the three contributions to \meff in 
Eq. (\ref{meffcompleteNH}) are relevant.
We consider the effect of cancellations 
between the three terms 
in the case of CP-invariance, in which 
there are four different 
neutrino CP-parity patterns. 
We will denote them as $+++$ ($+--$) 
if $\alpha_{21, 31} = 0~(\pi)$, and $++-$ ($+-+$) 
when $\alpha_{21}= 0~(\pi)$ while $\alpha_{31}=\pi~(0)$.
The prediction in the case of CP-violation will lie
within the ones obtained for CP-conservation.
Obviously, if both $0 \ltap \alpha_{21} \ltap \pi/2$ 
and $0 \ltap \alpha_{31} \ltap \pi/2$, there will be 
no mutual compensation between the three terms in 
Eq. (\ref{meffcompleteNH}) and we would have 
$\meff \gtap 3\times 10^{-3}$ eV.

    For each CP-parity pattern,
we analyse what are the values of 
$m_1$  and $\sin^2\theta_{13}$ 
which would guarantee $\meff \geq 1$~meV or, conversely, 
which would be implied by a negative 
result for a search of neutrinoless
double beta decay with a sensitivity of 1 meV, 
in the hypothesis of Majorana neutrinos.
The effective Majorana mass 
parameter would be predicted to be
smaller than 1 meV, if a sufficient 
cancellation between the three terms  in the r.h.s. 
of Eq.~(\ref{meffcompleteNH}) takes place.

   Here we use $\mu=1$~meV as a reference value 
for $\meff$, but similar results can be obtained 
for other values of $\meffref$ in the few~meV range~\footnote{Let us note that
a similar analysis
for $\mu=0$ was performed in Ref.~\cite{PPW}.}.
The central value of $m_1$ can be found
by solving Eq.~(\ref{meffcompleteNH}) 
with $\meff = 1$~meV,
while the error on $m_1$ is obtained by 
propagating the errors on the 
oscillation parameters:
\begin{equation}
\sigma(m_1)   =  
\left( \displaystyle\frac{\partial \meff}{\partial m_1} 
\right)^{-1} \! \! \! \! \sigma(\meff) 
 \simeq   \frac{\sigma(\meff)}{\cos^2 \theta_\odot  
\pm \displaystyle\frac{m_1 \sin^2 {\theta_\odot} }
{\sqrt{m_1^2 + \deltasol}}}~.
\end{equation}
%
The degree of cancellation between the three terms in 
$\meff$  depends on the neutrino CP-parity pattern.
The results for $m_1$ for the different 
CP-parity patterns are presented in Fig.~6
for three values of $\theta_\odot$,
$\sin^2 \theta_\odot = 0.26, 0.32, 0.40$,
using the prospective relative errors of 2\%, 2\% and 4\% for \deltaatm, \deltasol
and $\sin^2 \theta_\odot$ and the absolute error of 0.006 on $\sin^2 \theta_{13}$.
%
\begin{center}
 \begin{figure}
 \label{figm1}
\begin{center}
\begin{tabular}{c}
\includegraphics*[height=0.3\textheight, width=0.7\textwidth]{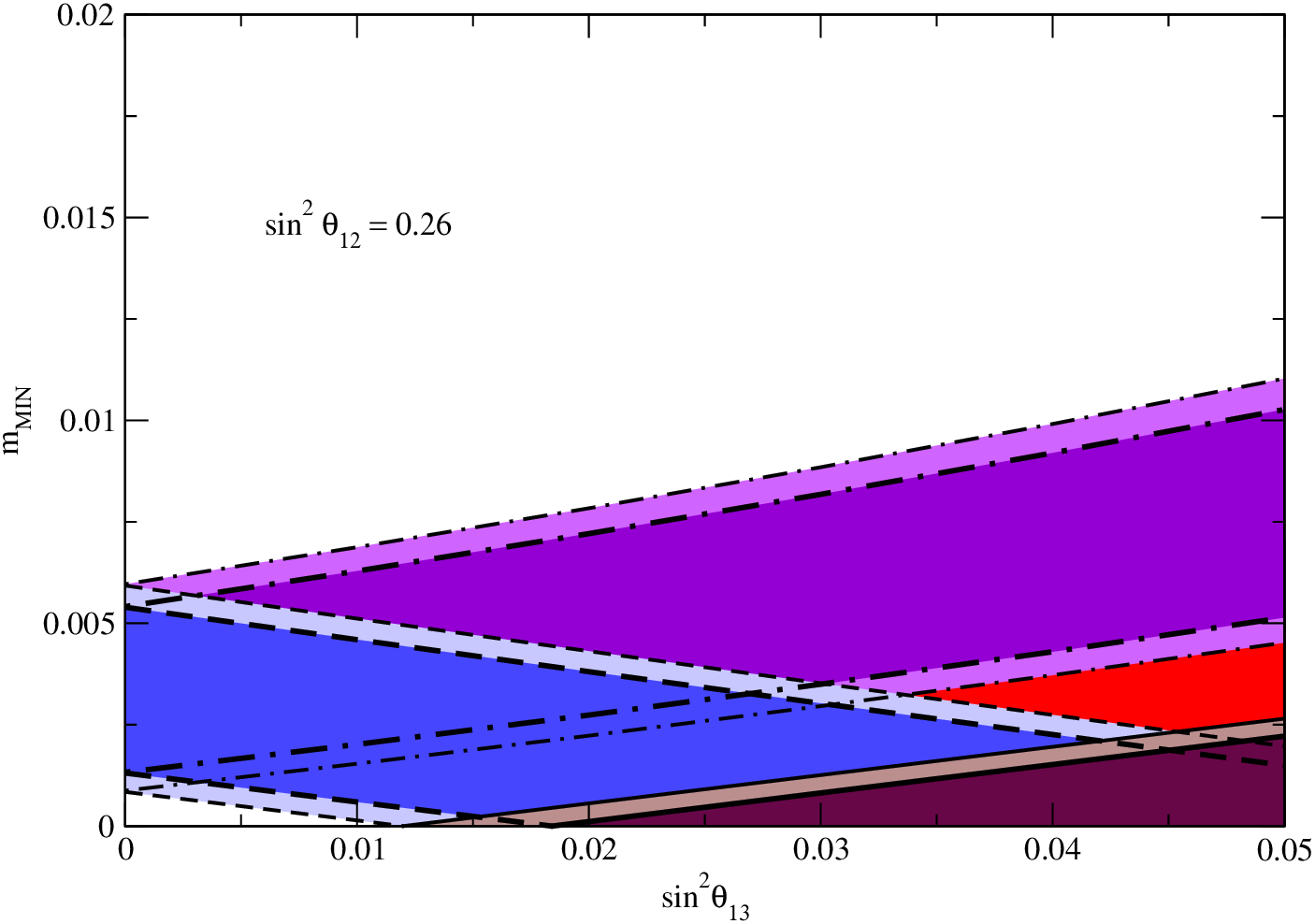} \\ 
\includegraphics*[height=0.3\textheight, width=0.7\textwidth]{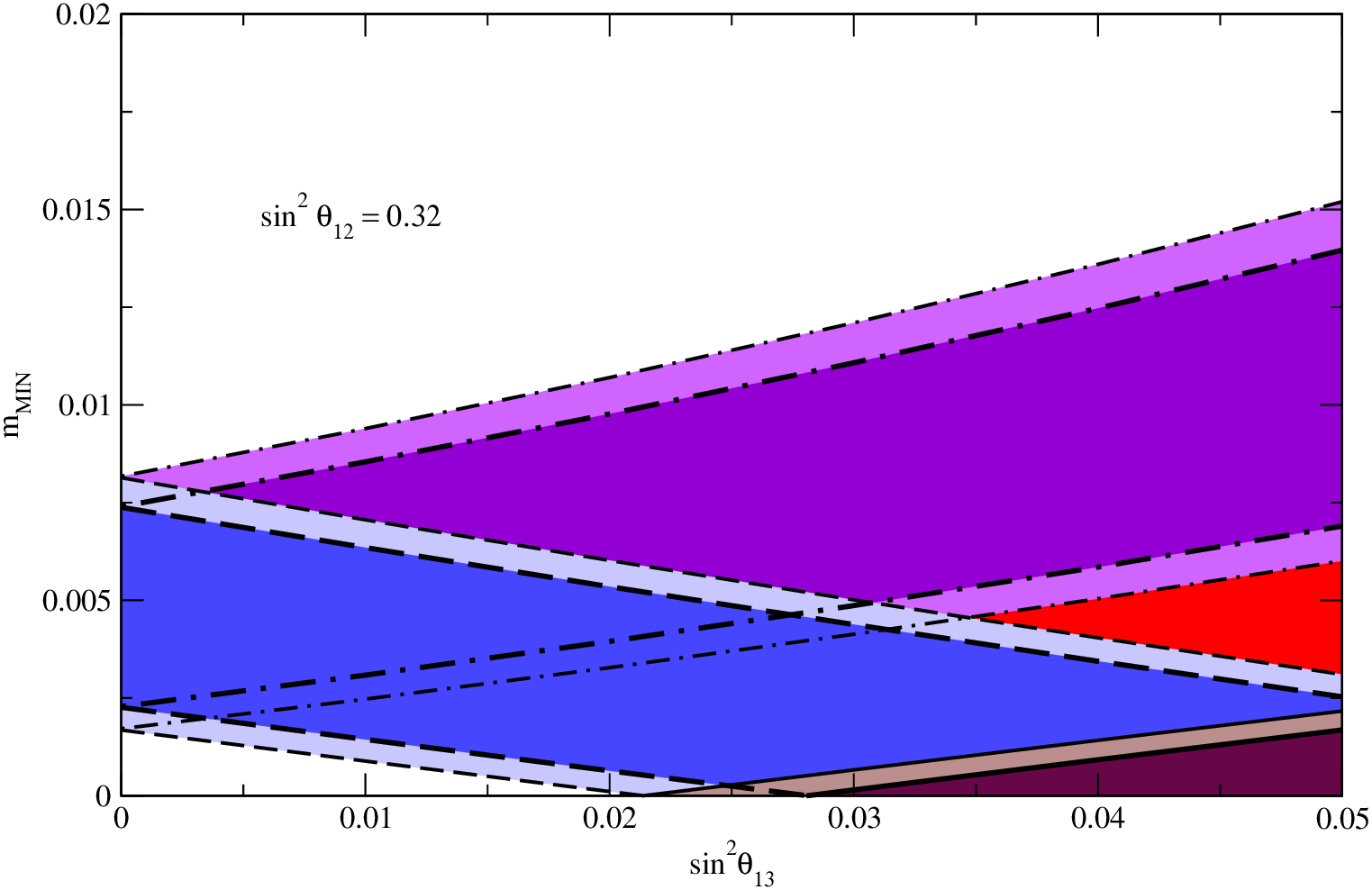} \\ 
\includegraphics*[height=0.3\textheight, width=0.7\textwidth]{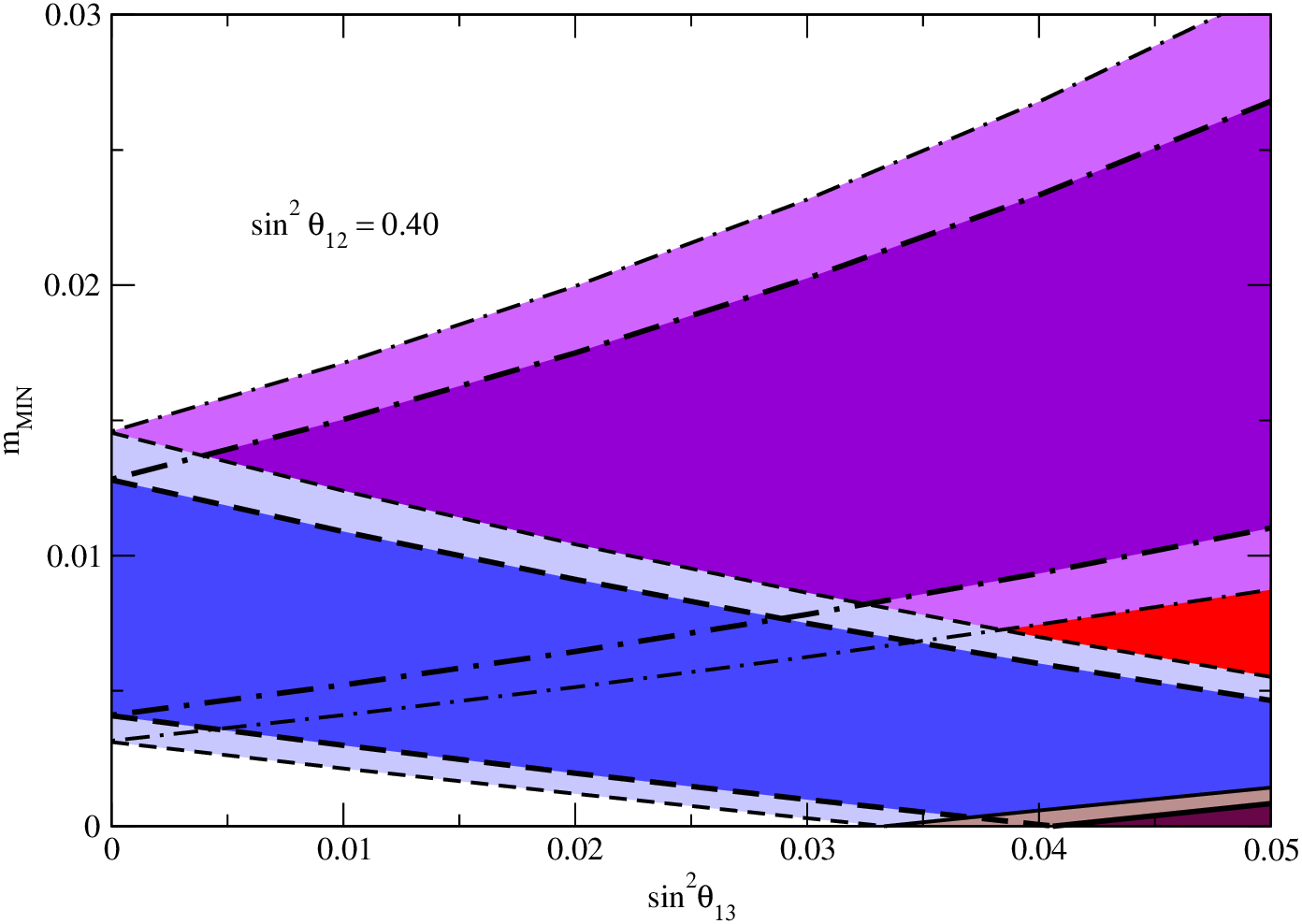} \\ 
\end{tabular}
\caption{The grey (colour online) regions denote the ranges of \protect\mmin~
for which $\protect\meff < 1$~meV and are delimited 
by thick (thin) lines at 1~(2)~$\sigma$.
The CP-conserving patterns are indicated by
i) solid lines for the case $++-$, ii) dashed lines for 
the $+-+$ one, and iii) dashed-dotted lines for $+--$.
The red triangular region requires CP-violation.
The present best fit values for \protect\deltasol and 
\protect\deltaatm~are used.
}
\end{center}
\end{figure}
\end{center}
%

We can understand the results in Fig.~6
by performing a simplified analysis
neglecting $\sigma(m_1)$. We study each 
CP-parity pattern separately. In the following,
we will use the present best fit values of \deltasol, \deltaatm~and
$\sin^2 \theta_\odot$, unless otherwise indicated.
\begin{itemize}
\item
For the CP-parity pattern $(+++)$, 
no cancellation takes place and 
we will have $\meff  \gtap 2.5$~meV for any allowed 
value of $\theta_{13}$ and $\theta_\odot$.
A negative search for neutrinoless double beta decay 
with a sensitivity of few meV, such that 
 $\meff_0  < \meff_+ - n \sigma(\meff)$, 
where $\meff_0$ is the experimentally determined 
value of $\meff \!\!$ and $\meff_+$ 
corresponds to NH spectrum and $\eta_{32} = +1$ 
(see Section 3.1), would strongly disfavour 
(if not rule out) this possibility.
\item
In the $(++-)$ case, 
a significant cancellation can take place 
only if the atmospheric term, 
$\sqrt{\deltaatm} \sin^2 \theta_{13}$, 
is of the same order as the sum of the first two terms in 
the r.h.s. of Eq.~(\ref{meffcompleteNH}). We can have 
$\meff \geq \meffref = 1$ meV for given $m_1$
provided  $\sin^2\theta_{13}$ satisfies
\begin{equation}
\sin^2 \theta_{13} \leq
\frac{m_1\cos^2 \theta_\odot + \sqrt{m_1^2 + \deltasol}
\sin^2 \theta_\odot -\meffref}{\sqrt{m_1^2 + \deltaatm}}\,.
\label{stheta13cond}
\end{equation}
%
The above inequality is always  fulfilled for 
$m_1 \gtap \sqrt{\deltasol}$. For 
$m_1 \ll \sqrt{\deltasol}$, this condition becomes
$\sin^2 \theta_{13} \leq (\sin^2 \theta_{13})_0$ with
\begin{equation}
(\sin^2 \theta_{13})_0 =
\sqrt{\displaystyle\frac{\deltasol}{\deltaatm}} \sin^2 \theta_\odot - 
\displaystyle\frac{\meffref}{\sqrt{\deltaatm}}\,.
\label{stheta13cond2}
\end{equation}
%
It is satisfied 
for the best fit values of 
$\deltasol\!\!$, $\sin^2 \theta_\odot$ and 
$\deltaatm$, while if one uses the 3$\sigma$ 
allowed ranges 
of these parameters, the inequality implies 
$\sin^2\theta_{13} \ltap 0.026, 0.036, 0.051$ for $\sin^2 \theta_\odot = 0.26, 0.32, 0.40$.
These values of $\sin^2\theta_{13} $ are close to the present 
3$\sigma$ upper bound.
In summary,  for values of $\sin^2 \theta_{13}\leq (\sin^2 \theta_{13})_0$,
\meff is guaranteed to be larger than 1~meV for any $m_1$.

For given $\sin^2 \theta_{13}>(\sin^2 \theta_{13})_0$,   
we will have $\meff \geq \meffref = 1$ meV 
 if $m_1$ 
satisfies $m_1 \geq (m^A_1)_-$,
where 
\begin{eqnarray}
(m^A_1)_{\pm}  &=& \frac{1}{\cos 2 \theta_\odot}\,
\Big [ \Big(\sqrt{\deltaatm } \sin^2 \theta_{13}
+ \meffref  \Big) \cos^2 \!\theta_\odot\,~
\nonumber \\[0.25cm]
&\pm &~ 
\sin^2 \!\theta_\odot \sqrt{\,\Big(\sqrt{\deltaatm} 
\sin^2 \!\theta_{13}+ \meffref \Big)^2 
+  \deltasol \cos \!2\theta_\odot }\, \Big ]\,.
\label{m1PH1}
\end{eqnarray}
%
In deriving Eq. (\ref{m1PH1}) we have neglected $m_1^2$ 
with respect to $\deltaatm$ and 
have taken $\cos^2 \theta_{13} \simeq 1$.
The lower bound $(m^A_1)_-$
of $m_1$ in Eq. (\ref{m1PH1}) 
increases with $\sin^2 \theta_{13}$, 
but is rather small:
for $\meffref = 10^{-3}$ eV, 
$\sin^2 \theta_{13} = 0.05$ and best fit 
values of the other relevant oscillation 
parameters we get $(m^A_1)_- \cong 0.9 \times 10^{-3}$ eV. 
\item
For the CP-parity pattern $(+-+)$, 
a partial cancellation can take place 
between the first and the second terms in 
Eq.~(\ref{meffcompleteNH}); the cancellation 
would be significant only if $m_1 \sim$~few~meV.
The second term in Eq.~(\ref{meffcompleteNH}) 
would dominate and we would have
$\meff \geq \meffref = 10^{-3}$ eV only if 
$m_1$ and $\sin^2 \theta_{13}$ are 
sufficiently small, more precisely, 
if $0 \leq m_1 \leq (- (m^A_1)_-)$ and
$\sin^2 \theta_{13} \leq (\sin^2 \theta_{13})_0$,
where $(m^A_1)_-$ and $(\sin^2 \theta_{13})_0$ are
given in Eqs. (\ref{m1PH1}) and (\ref{stheta13cond2}), 
respectively. 

\hskip 1.0cm The sum of the first and third terms 
in Eq. (\ref{meffcompleteNH}) will 
dominate and will lead to 
$\meff \geq \meffref = 10^{-3}$ eV for $m_1 \geq (m^B_1)_+$, where
 \begin{eqnarray}
(m^B_1)_\pm & = & \frac{1}{\cos 2 \theta_\odot}\,
\Big [ \Big(\, \meffref - \sqrt{\! \deltaatm} \,  \sin^2\theta_{13}\,\Big)\, 
\cos^2 \!\theta_\odot  \nonumber \\ [0.25cm]
& \pm & \sin^2 \!\theta_\odot \sqrt{\Big(\, \meffref - \sqrt{\! \deltaatm }  \,
\sin^2\theta_{13}\,\Big)^2 + \deltasol\! \cos \!2\theta_\odot }\, \Big ]\,,
\label{m1PH2}
 \end{eqnarray}
%
provided 
\begin{equation}
\sin^2 \theta_{13} \leq
\sqrt{\displaystyle\frac{\deltasol}{\deltaatm}} \sin^2 \theta_\odot + 
\displaystyle\frac{\meffref}{\sqrt{\deltaatm}}\,.
\label{stheta13cond3}
\end{equation}
%
Given the experimental 3$\sigma$ upper bound 
$\sin^2 \theta_{13} < 0.05$, the second inequality 
is always satisfied for $\meffref = 10^{-3}$ eV.
For $\sin^2 \theta_{13} = 0~(0.02)$ we get from 
Eq.~(\ref{m1PH2}): $m_1 \gtap 6.6~(4.7)\times 10^{-3}$ eV.
\item
Finally, consider the case $(+--)$.
As the second and third term in the r.h.s. 
of Eq.~(\ref{meffcompleteNH})
are summed constructively,
a strong cancellation in \meff can happen only for 
sufficiently large values of $m_1$. 
We get $\meff \geq \meffref = 10^{-3}$ eV for 
$0 \leq m_1 \leq (-(m_1^B)_-)$ 
and for $m_1 \geq (m^A_1)_+$, 
where $(m^A_1)_+$ is given in 
Eq. (\ref{m1PH1}). 
The maximal value of $m_1$ 
determined by Eq. (\ref{m1PH2}) can be rather large.
More specifically, we have
$-(m^B_1)_-= 2.8~(5.0)~[7.6]\times 10^{-3}$~eV
for $\sin^2 \theta_{13} = 0~(0.025)~[0.05]$.
For the minimal value of $m_1$
determined by the inequality $m_1 \geq (m^A_1)_+$, 
we get for $\sin^2 \theta_{13} = 0~(0.025)~[0.05]$:
$(m^A_1)_+ = 6.6~(9.3)~[12.1] \times 10^{-3}$~eV.
In the latter case the sum of the neutrino 
masses is limited from below 
by $(m_1 + m_2 + m_3) \gtap  6.8~(7.2)~[7.9]\times 10^{-2}$ eV.
Both $(-(m^B_1)_-)$ and
$(m^A_1)_+$  increase 
with $\theta_{13}$ and $\sin^2 \theta_\odot$.
\end{itemize}

 It follows from the preceding discussion that
if a future highly sensitive 
\betabeta-decay experiment does not find
a positive signal down to $\meff \sim 1$~meV,
Majorana neutrinos would still be allowed,
but the spectrum would be constrained
to be with normal ordering and
$m_1$ would be bound to be
smaller than $\sim 10^{-2}$ eV.
The CP-parity pattern (+++) 
will be strongly disfavored (if not ruled out)
as well.
If in addition it is found that 
$\sin^2 \theta_{13} \ltap 0.01$, 
i) the CP-parity pattern ($++-$) 
will also be disfavored, and
ii) $m_1$ would be constrained to 
lie in the interval 
$m_1 \sim (10^{-3} - 10^{-2})$ eV.
No other future neutrino 
experiment will have the capability
of constraining the lightest neutrino mass 
(and the absolute neutrino mass scale) 
in the meV range. Obviously, the above 
limits would hold only if 
massive neutrinos are Majorana particles. 
If the lightest neutrino has a mass 
in the interval $m_1 \sim (10^{-3} - 10^{-2})$ eV, 
this can have important effects 
on the generation of the baryon asymmetry of 
the Universe in the ``flavoured'' 
leptogenesis scenario of matter-antimatter 
asymmetry generation \cite{MPST07}.


\section{\betabeta-Decay in the Case of $\meff = 0$}


 In the present Section we shall discuss briefly 
the possible implications of having $\meff = 0$ 
for the process of $\betabeta-$decay. 
If $\meff = 0$ as a consequence of conservation 
of certain lepton charge, which could be, e.g. 
$L_e$, $L$, or $L' = L_e - L_{\mu} - L_{\tau}$, 
the $\betabeta-$decay will be strictly forbidden.
However, in the case of neutrino mass spectrum with 
normal ordering, one can have $\meff = 0$, 
as we have seen, as a consequence of an 
``accidental'' relation involving the neutrino 
masses, the solar neutrino and CHOOZ mixing angles 
and the Majorana phase(s) in $\pmns$.
For the spectrum of the normal hierarchical 
type, the relation of interest is given by 
Eq. (\ref{NHmeff0}), while if $\sin^2\theta_{13}$ 
is negligibly small it is shown in 
Eq. (\ref{NHth130meff0}). 
None of the two relations can be  
directly associated with a 
symmetry which forbids 
$\betabeta-$decay. Thus, 
if $\meff = 0$ is a consequence 
of Eq. (\ref{NHmeff0}) 
or Eq. (\ref{NHth130meff0}), 
$\betabeta-$decay will 
still be allowed. In 
what follows we will estimate 
the non-zero contribution to the 
$\betabeta-$decay amplitude $A\betabeta$
due to the exchange of 
the light massive Majorana neutrinos 
$\nu_j$ in the case when 
$\meff = 0$ and there is no symmetry 
forbidding the decay.

  Suppose that neutrino masses and 
mixing arise due to the Majorana mass term 
of the three flavour neutrinos: 
\beq 
\mathcal{L}^{M}(x) =  
- ~\frac{1}{2}~m_{ll'}\, \overline{\nu^{c}}_{lR}~\nu_{l'L}
+ h.c.
\label{Majm1}
\eeq
%
where $\nu^{c}_{lR} = C (\bar{\nu}_{lL})^{\rm T}$, 
$l=e,\mu,\tau$, $C$ being the charge conjugated matrix. 
We have $m_{ll'} = m_{l'l}$, $l,l'=e,\mu,\tau$ 
(see, e.g. \cite{BiPet87}). The mass term 
in Eq. (\ref{Majm1}) is diagonalised using the 
congruent transformation: 
$m = U^*\, m^{d}\, U^{\dagger}$, where 
$m^d = diag(m_1,m_2,m_3)$ is a diagonal matrix 
formed by the masses of the Majorana neutrinos 
$\nu_j$ and $U$ is the PMNS matrix 
\footnote{We work in the basis in which 
the charged lepton mass matrix is diagonal.}.
The effective Majorana mass 
$\mefff$ arises in $A\betabeta$ from 
the virtual neutrino propagator (see, e.g. \cite{BiPet87}):
\beq 
\mathcal{P} = \sum_{j} U^2_{ej} \, \frac{m_j}{q^2 - m^2_j} = 
P_1 + P_3 + P_5 + ...\,,
\label{MajP1}
\eeq
%
where 
\begin{eqnarray}
P_1 = \frac{1}{q^2}\, \sum_{j} U^2_{ej} \, m_j = 
\frac{1}{q^2}\, \mefff \,,\\[0.25cm]  
P_3 =  \frac{1}{q^2}\, \sum_{j} U^2_{ej}\,m_j\, \frac{m^2_j}{q^2}\,,~ 
{\rm etc.}
\label{MajP2}
\end{eqnarray}
%
Here $q$ is the momentum of the virtual 
neutrino and we have used 
the fact that $m^2_j \ll |q^2|$. Typically 
one has for the 
average momentum of the virtual neutrino in 
$\betabeta-$decay (see, e.g. \cite{Haxton91}): 
$|q^2|\sim (10~{\rm MeV})^2$. As a consequence, 
the following inequalities hold
$|P_{2n+1}| \ll |P_1|$, $n=1,2,...$. 
Usually the terms $P_3,~P_5$, etc. are neglected 
in the expression for $\mathcal{P}$. The dominant term 
$P_1 \propto \mefff$, which leads to 
$A\betabeta \propto \mefff$. The $q^{-2}$ factor in 
$P_1$ gives rise to a Coulomb-like potential 
of interaction between the nucleons 
exchanging the virtual neutrino in the 
nucleus undergoing $\betabeta-$decay.

 Assume now that  $\meff = 0$. In this case $P_1 = 0$ and the
dominant term in the expression for  $\mathcal{P}$, 
Eq. (\ref{MajP1}), will be $P_3$. If $\meff = 0$ 
is not a consequence of a conservation 
of some lepton charge, we will have 
$P_3 \neq 0$ and $A\betabeta \neq 0$, in general.
However, unless the $\betabeta-$decay amplitude 
receives contributions from mechanisms other than 
the exchange of the light Majorana neutrinos $\nu_j$,
the  $\betabeta-$decay rate will be extremely
strongly suppressed due to the fact that \cite{Haxton91}
$m_j^2/|q^2| < 10^{-14}$, where we have used 
$m_j < 1$ eV. Although allowed, $\betabeta-$decay  
will be practically unobservable 
if the $P_3$ term in $\mathcal{P}$
gives the dominant contribution in $A\betabeta$.

  It is well-known (see e.g. \cite{BiPet87}) 
that $\meff = |m_{ee}|$,  
where $m_{ee}$ is the $ee-$element of the 
Majorana mass matrix $m$ of neutrinos, Eq. (\ref{Majm1}).       
If  $\meff = |m_{ee}| = 0$, 
the term $\overline{\nu^{c}}_{eR}~\nu_{eL}$ 
will effectively be ``regenerated'' 
at higher orders from the other 
terms in $\mathcal{L}^{M}(x)$, Eq. (\ref{Majm1}).
The exchange of virtual $\nu_e$ 
mediated by this term will lead to 
$\betabeta-$decay. If we treat 
$\mathcal{L}^{M}(x)$ as an ``interaction'' term 
\footnote{In this case $\nu_{lL}(x)$ 
should be considered as zero mass fermion
fields having the standard zero mass 
fermion propagator.} 
and use perturbation theory, 
the virtual neutrino propagator 
in the $\betabeta-$decay amplitude 
will have, to leading order in the parameters 
$m_{ll'}$, the following form:   
\beq 
\mathcal{P} = \frac{1}{q^2}\, \frac{\tilde{m}^*}{q^2} + ...\,,
\label{MajP3}
\eeq
%
where 
\beq 
\tilde{m} = m_{e\mu}\, m^*_{\mu \tau}\, m_{\tau e} + 
m_{e\mu}\, m^*_{\mu \mu}\, m_{\mu e} + 
m_{e\tau}\, m^*_{\tau \mu}\, m_{\mu e} + 
m_{e\tau}\, m^*_{\tau \tau}\, m_{\tau e}\,.
\label{tm1}
\eeq
%
It follows from the expression for the 
mass parameter $\tilde{m}$ that if $m_{ee} = 0$, we 
will have $\tilde{m} = 0$ in the following 
cases \cite{STPPD82,LeungP84,HirschKov06}:\\
i) $m_{e\mu} = m_{e \tau} = 0$, 
ii) $m_{e\mu} = m_{\tau \tau} = 0$, 
iii) $m_{e\tau} = m_{\mu \mu} = 0$, 
iv)  $m_{\tau \mu} = m_{\mu \mu} = m_{\tau \tau} = 0$.
It is easy to see that the four cases
in which $\tilde{m} = 0$ correspond 
to the conservation 
of the following lepton charges \cite{LeungP84}:
i) $L_e$, 
ii) $L_{e} - L_{\tau}$, 
iii) $L_{e} - L_{\mu}$, 
iv)  $L_{e} - L_{\mu} - L_{\tau}$. 
In all these four cases the $\betabeta-$decay 
is strictly forbidden. However, 
all four cases are ruled out 
by the existing neutrino oscillation 
data (see, e.g. \cite{HirschKov06,WRode06}).
Thus, we can conclude that $\tilde{m} \neq 0$ 
and therefore $A\betabeta \neq 0$.

  How large can the mass parameter $\tilde{m}$
be? Using the relation $m = U^*\, m^{d}\, U^{\dagger}$
and assuming that $m_{ee} = \mefff = 0$,
it is not difficult to show that 
\beq 
\tilde{m}^* =  \sum_{j} U^2_{ej}\, m^3_j\,. 
\label{tm2}
\eeq
%
Thus, we recover the result obtained earlier 
by expanding the 
massive Majorana neutrino propagators 
in power series of $m^2_j/q^2$: 
\beq 
\mathcal{P} = \frac{1}{q^2}\, \frac{\tilde{m}^*}{q^2} + ... = 
P_3 + ...
\label{MajP4}
\eeq
%
where $P_3$ is given in Eq. (\ref{MajP2}).
Therefore the $\betabeta-$decay will be 
extremely strongly suppressed 
if $m_{ee} = 0$ and $A\betabeta \neq 0$
is generated at higher order by the 
Majorana mass term, Eq.~(\ref{Majm1}).  


\section{Conclusions}


  Present and future searches for neutrinoless 
double beta decay aim at probing lepton number 
violation and the Majorana nature of neutrinos with 
remarkable precision.
A wide experimental program 
is currently under discussion. 
Experiments with a sensitivity to the
effective Majorana mass parameter, 
$\meff\!\!$, down to $\sim$~(50 - 10) meV  
are in a stage of preparation or planning 
and will take place in the future.
These experiments will provide 
valuable information on 
the neutrino masses and the nature 
of massive neutrinos. 

  If future \betabeta-decay experiments 
show that $\meff < 0.01$ eV, both the IH and
the QD spectrum will be ruled out for massive 
Majorana neutrinos. If in addition it is 
established in neutrino oscillation 
experiments that the neutrino mass spectrum is 
with {\it inverted ordering}, i.e. that $\deltaatm < 0$,
the absence of a signal in neutrinoless double beta 
decay experiments sensitive to $\meff \sim 10$ meV
would be a strong indication that the massive 
neutrinos $\nu_j$ are Dirac fermions.
At the same time 
the alternative explanation 
based on the assumptions that the 
massive neutrinos $\nu_j$ are 
Majorana particles but there are 
additional contributions to the 
\betabeta-decay amplitude which 
interfere destructively 
with that due to the exchange of 
light massive Majorana neutrinos, 
would also be possible.
However, if $\deltaatm$ is determined 
to be positive in neutrino oscillation 
experiments, the upper limit $\meff < 0.01$ eV 
would be perfectly compatible with 
massive Majorana neutrinos
possessing {\it normal hierarchical} 
mass spectrum, or mass spectrum with 
{\it normal ordering but partial hierarchy}, 
and the quest for $\meff$ would still be open.
Under such circumstances
the next frontier in the searches for 
$\betabeta-$decay would most probably 
correspond to values of $\meff \sim 0.001$ eV.

  Taking $\meff = 0.001$ eV as a reference value,
we have investigated in the present article
the conditions under which $\meff$ in the 
case of neutrino mass spectrum with normal 
ordering would satisfy $\meff \gtap 0.001$ eV.
We have considered the specific cases of 
i) normal hierarchical neutrino mass spectrum, 
ii) of relatively small value of 
the CHOOZ angle $\theta_{13}$, as well as 
iii) the general case of spectrum 
with normal ordering, partial hierarchy 
and a value of $\theta_{13}$ close to the 
existing upper limit. We have 
derived the ranges of the lightest 
neutrino mass $m_1$ and/or of 
$\sin^2\theta_{13}$, for which 
$\meff \gtap 0.001$ eV, and have 
discussed some related 
phenomenological implications. 
We took into account the uncertainties in the 
predicted value of $\meff$ due to the 
uncertainties in the measured values 
of the input neutrino oscillation 
parameters $\dmsol$, $\deltaatm$ and
$\sin^2\theta_{\odot}$.
For the latter we have used the 
following prospective 1$\sigma$ errors:
2\%, 2\% and  4\%, respectively.

 In the present analysis we did not 
include the possible effects of the 
uncertainty related to the imprecise knowledge of the 
$\betabeta-$decay nuclear matrix elements.
We hope (perhaps optimistically) 
that by the time it will become 
clear whether the searches for 
$\betabeta-$decay will require 
a sensitivity to values of 
$\meff < 0.01$ eV, the problem of 
sufficiently precise calculation of 
the $\betabeta-$decay nuclear matrix 
elements  will be resolved.

  We have found that in the case of 
normal hierarchical (NH) neutrino 
mass spectrum we get $\meff \gtap 0.001$ eV
for $\sin^2\theta_{13} \ltap (0.01-0.02)$
and any value of the relevant Majorana phase 
(difference) $\alpha_{32}$, provided 
the currently determined best fit values 
of the solar and atmospheric 
neutrino oscillation parameters 
$\deltasol$, $\deltaatm$ 
and especially of $\sin^2 \theta_\odot$,
will not change considerably
in the future high precision 
measurements (Fig. 3). 
For  $0 \leq \alpha_{32} \leq \pi/2$
one has $\meff \gtap 2.0\times 10^{-3}$ eV
for any $\sin^2\theta_{13}$, while if 
$\pi/2 < \alpha_{32} \leq 5\pi/6$,
we get $\meff \gtap 10^{-3}$ eV 
for any $\sin^2\theta_{13}$
allowed at 3$\sigma$ by the existing data.
Values of $\alpha_{32}\neq 0$  in the indicated 
ranges are required for the generation 
of the baryon asymmetry of the Universe 
in the ``flavoured'' leptogenesis scenario, 
in which the requisite CP-violation is provided 
exclusively by the Majorana phase (difference) 
 $\alpha_{32}$ \cite{PPRio106}.

 We have investigated also the case  
when $\sin^2\theta_{13}$ has a rather small value,
$\sin^2\theta_{13} \ltap 3\times 10^{-3}$,
but the neutrino mass spectrum is not hierarchical.
We have found that in this case 
one has $\meff \geq 10^{-3}$ eV for any 
value of the relevant Majorana phase $\alpha_{21}$
if the sum of neutrino masses satisfies 
$m_1 + m_2 + m_3 \gtap 7\times 10^{-2}$ eV.

  In the general case of neutrino mass spectrum 
with partial hierarchy (i.e. non-negligible 
lightest neutrino mass $m_1$) 
and sufficiently large $\sin^2\theta_{13}$,
one finds $\meff \geq 10^{-3}$ eV typically 
for $m_1 \ltap {\rm few}\times 10^{-3}$ eV and 
$m_1 \gtap 10^{-2}$ eV (Fig. 6). In the second case 
the sum of neutrino masses
satisfies $m_1 + m_2 + m_3 \gtap 7.4\times 10^{-2}$ eV.
If a future highly sensitive 
\betabeta-decay experiment does not find
a positive signal corresponding to $\meff \geq 1$~meV,
Majorana neutrinos would still be allowed,
but the spectrum would be constrained
to be with normal ordering and
$m_1$ to be
smaller than $\sim 10^{-2}$ eV.
The CP-parity pattern (+++) 
will be strongly disfavored (if not ruled out)
as well. If in addition it is found that 
$\sin^2 \theta_{13} \ltap 0.01$, 
$m_1$ would be constrained
to lie in the interval 
$m_1 \sim (10^{-2} - 10^{-3})$ eV
(for $\sin^2 \theta_\odot \sim 0.32$),
and the CP-parity pattern ($++-$) 
will also be disfavored.
No other future neutrino 
experiment, foreseeable at present, 
will have the capability
of constraining the lightest neutrino mass 
(and the absolute neutrino mass scale) 
in the meV range. Obviously, the above 
constraints would hold only if 
massive neutrinos are Majorana particles. 
If the lightest neutrino has a mass 
in the interval $m_1 \sim (10^{-3} - 10^{-2})$ eV, 
this can have important effects 
on the generation of the baryon asymmetry of 
the Universe in the ``flavoured'' 
leptogenesis scenario of matter-antimatter 
asymmetry generation \cite{MPST07}.

  We have provided also an estimate 
of $\meff$ when the three neutrino masses and the neutrino 
mixing originate from neutrino mass term of Majorana type 
for the (left-handed) flavour neutrinos and 
$\sum^{3}_{j} m_j U^2_{ej} = 0$, but 
there does not exist a symmetry which forbids 
the $\betabeta$-decay. Our results show 
that, although in this case the $\betabeta$-decay 
will be allowed, the corresponding 
effective Majorana mass parameter
is determined by $\sum^{3}_{j} m^3_j U^2_{ej}/q^2$, 
where $q$ is the momentum of the virtual 
Majorana neutrino. For the average momentum of the 
virtual neutrino in $\betabeta$-decay 
one typically has (see, e.g. \cite{Haxton91}): 
$|q^2|\sim (10~{\rm MeV})^2$. 
As a consequence the contribution to the 
$\betabeta$-decay amplitude $A\betabeta$
due to the light Majorana neutrino exchange
will be strongly suppressed: 
$\meff \ll 10^{-3}$ eV.
Thus, if $\sum^{3}_{j} m_j U^2_{ej} = 0$
and $\betabeta$-decay is observed in 
an experiment with sensitivity to 
$\meff \sim 10^{-3}$ eV, that would imply 
the existence of contributions 
to $A\betabeta$ due to mechanism(s) 
other than the three light Majorana neutrino 
exchange.

\vspace{-0.4cm}
\section{Acknowledgments}
\vspace{-0.3cm}

Part of this work was done at the Aspen Center for Physics
during the 2007 Summer Program on Neutrino Physics.
We acknowledge support by the European Network of Theoretical 
Astroparticle Physics  ILIAS/N6 under the contract RII3-CT-2004-506222. 
This work was also supported in part by the INFN
program on ``Astroparticle Physics'' as well as by the 
Italian MIUR (PRIN and Internazionalizzazione Programs)
and the Yukawa Institute of Theoretical Physics 
(YITP), Kyoto, Japan, within the joint SISSA--YITP 
research project on ``Fundamental Interactions 
and the Early Universe'' (S.T.P.).

\end{document}